\documentclass[pra,showpacs,groupedaddress,amssymb,twocolumn,nofootinbib,longbibliography,floatfix,superscriptaddress]{revtex4-1}
\usepackage{graphicx,amsmath,dsfont}
\usepackage[dvipsnames]{xcolor}
\usepackage{animate}
\usepackage{tabularx}
\usepackage{enumerate}
\usepackage{float}
\usepackage{comment}
\usepackage{multirow}
\usepackage{physics}
\usepackage{algorithm}
\usepackage{algpseudocode}
\usepackage{color}
\usepackage{amsthm}
\usepackage{amsmath,amssymb,amsbsy}
\usepackage{calrsfs}
\usepackage[english]{babel}
\usepackage{xcolor}
\usepackage{ragged2e}
\usepackage{orcidlink}
\usepackage[capitalize]{cleveref}

\newcounter{protocol}

\pagecolor{white}

\newcommand{\hads}[1]{\hat{a}_{#1}^\dagger}

\renewcommand{\Im}{\mathrm{Im}}

\newcommand{\expect}[1]{\left\langle#1\right\rangle}
\newcommand{\eea}{\end{eqnarray}}
\newcommand{\bea}{\begin{eqnarray}}
\newcommand{\ee}{\end{equation}}
\newcommand{\be}{\begin{equation}}

    \newcommand{\id}{\mathds{1}}

\begin{document}

\title{Partially-Blind Single-Qubit Classification over a Prototype Hybrid Quantum Network}

\author{Matteo Pasini \orcidlink{https://orcid.org/0009-0005-1358-7896}}
\email{matteo.pasini@icfo.eu}
\thanks{These authors contributed equally to this work.}

\affiliation{ICFO—Institut de Ciencies Fotoniques, The Barcelona Institute of Science and Technology, Castelldefels (Barcelona) 08860, Spain}
\author{Tzula Benjamin Propp \orcidlink{https://orcid.org/0000-0002-7748-1434}}
\email{propp@physics.leidenuniv.nl}
\thanks{These authors contributed equally to this work.}

\affiliation{LION—Leiden Institute of Physics, Leiden University, 2333 CA Leiden, The Netherlands}
\author{Janice van Dam}
\affiliation{QuTech, Delft University of Technology, Delft, The Netherlands}
\affiliation{Kavli Institute of Nanoscience, Delft University of Technology , Delft, The Netherlands}
\author{Garazi Muguruza Lasa \orcidlink{https://orcid.org/0009-0006-4627-9520}}
\affiliation{SURF—Collaborating University Computing Facilities, 3511EP Utrecht, The Netherlands}
\author{Alexandre Wanick}
\affiliation{NITeQ—Interdisciplinary Nucleus for Quantum Technologies, PUC-Rio—Pontifical Catholic University of Rio de Janeiro, 22451-900 Rio de Janeiro, Brazil}
\author{Hugues de Riedmatten \orcidlink{https://orcid.org/0000-0002-4418-0723}}
\affiliation{ICFO—Institut de Ciencies Fotoniques, The Barcelona Institute of Science and Technology, Castelldefels (Barcelona) 08860, Spain}
\affiliation{ICREA—Institucio Catalana de Recerca i Estudis Avançats, 08015 Barcelona, Spain}
\author{Gustavo C. do Amaral \orcidlink{https://orcid.org/0000-0001-6071-7458}}
\affiliation{NITeQ—Interdisciplinary Nucleus for Quantum Technologies, PUC-Rio—Pontifical Catholic University of Rio de Janeiro, 22451-900 Rio de Janeiro, Brazil}
\affiliation{Quantum Technology Department, TNO—The Netherlands Organization for Applied Scientific Research, 2628CK Delft, The Netherlands}

\begin{abstract}
In the NISQ era, there is a need for resource-efficient proof-of-principle experiments that can be built up to genuine utility. Single-qubit classifiers (SQCs) are small-scale hybrid quantum-classical machines capable of performing a basic machine learning task: classifying data. In principle, these can be scaled up to many-qubit quantum classifiers capable of quantum computational advantage. Another type of quantum advantage is enabled by blind quantum computation (BQC), wherein a client may run delegated quantum computations on an untrusted server with information-theoretic security. In this paper, we develop a framework and propose a prototype experiment for a SQC where it is known to the server that a classification is being performed, but the data and outcome stay hidden, i.e., it performs partially-blind SQC (PB-SQC). This can be integrated into a quantum network to deliver quantum-secured classifications to remote clients; we study this for a heterogeneous quantum network link in which entanglement is shared between a server and a client equipped with a multiplexed solid-state quantum memory using entanglement swapping. The framework we develop for PB-SQC on this setup is tested in a simulation with realistic hardware parameters on a real-world credit card transaction fraud database with classification outcomes approaching those of its equivalent classical deep-belief network. In addition, we show how a two-qubit classifier (TQC) instead of a SQC enables verification of the computation. These results pave the way towards a short- to mid-term quantum network offering use-case-ready quantum applications. 
\end{abstract} 

\maketitle

\section{Introduction}

\noindent
Distributing entanglement between quantum devices in a network enables applications that can provide advantages over their purely classical counterparts, as well as radically novel technologies, such as quantum position verification \cite{kanneworff2025towards}, quantum time transfer \cite{dai2020towards}, different flavors of quantum key distribution \cite{jouguet2013experimental, lucamarini2018overcoming, ferreira2013proof, rubenok2013real, lu2026device}, quantum clock synchronization \cite{jozsa2000quantum}, and distributed quantum computing \cite{wehner2018quantum} and sensing \cite{stas2026entanglement}. One outstanding feature of quantum networks is the possibility to provide information-theoretically secure privacy for server-delegated quantum computations \cite{broadbent2009universal}. This application is compelling in a technological context where handling computations involving a significant amount of sensitive data is often delegated to classical providers and servers without a verifiable guarantee of privacy. The quantum-verified security of quantum networks enables clients to run advanced computational tasks on a quantum information processor controlled by a server while maintaining information-theoretic privacy. The set of protocols geared towards this application falls under the general framework of Blind Quantum Computing (BQC) \cite{broadbent2009universal}.

BQC protocols provide privacy for the structure, results, and underlying data at the basis of a computation, hiding it from a server that executes it, as well as any eavesdropper intercepting server-client communications and collaborating maliciously with the server; this is referred to as \textit{blindness}. Additionally, some implementations can provide \textit{verifiability}, that is, empowering the client to verify that the server is performing the computation correctly \cite{Gheorghiu2018}. In general, the strength of a BQC protocol lies in provable security, including in any real-life implementation up to tolerable noise levels. Blind quantum machine learning (B-QML) \cite{Li2024,Fleur2025BQML} protocols are a subset of BQC that implement hybrid quantum-classical or fully quantum versions of machine learning tasks blindly, in principle enabling quantum network users with limited quantum information processing capabilities to securely run complex QML tasks on an external quantum server. This is further motivated by the notion that, at least in the short-term, large quantum processors will likely be centralized and made available through a quantum network or internet \cite{wehner2018quantum}.

 Classification is a form of supervised machine learning, where a quantum speedup can be achieved even when using few qubits to create a quantum neural network via data-reuploading \cite{PrezSalinas2020}, trading a space overhead in a traditional quantum neural network for a time overhead. For a single qubit classifier (SQC) this overhead is exponential, precluding a quantum advantage in speed. However, a SQC is a quantum learning task that can be made blind via BQC, forming a rudimentary form of B-QML amenable to implementation on a near-term quantum network, as we propose in this work. 

Single- or multi-client accessibility to a centralized quantum server to provide B-QML, including blind classification, in a large-scale network relies on an infrastructure that enables efficient long-distance quantum links, with the possibility of generating entanglement beyond the metropolitan scale. At the same time, the ability of a client node to perform remote operations is necessary for successfully enabling BQC protocols \cite{broadbent2009universal}; remote state preparation (RSP) of server qubits is one such operation with varied implementations \cite{Pati2000,Bennett2001,Leung2003,alexandru2019,Xin2023,Chen2024,Drmota2024,vanDam2024,vanDam2025,Propp2026}, and is a cryptographic prerequisite for measurement-based implementations of BQC.

Quantum repeaters offer a way to efficiently distribute long-distance entanglement, overcoming photon loss in direct transmission, by performing rounds of entanglement swapping between intermediate segments aided by quantum memories \cite{bennett1993teleporting, askarani2021entanglement}. Experimental progress in the field led to the demonstration of entanglement distribution between remote quantum memories \cite{stolk2024metropolitan, lago2021telecom, yu2020entanglement}, the basis of elementary repeater links. Among the leading hardware platforms in this field are atomic ensembles \cite{maring2017photonic, wang2021cavity} and solid-state rare-earth ion-based memories connected using entangled photon pair sources \cite{simon2007quantum, afzelius2009multimode, businger2022non, askarani2021long}. One outstanding experimental challenge toward a fully functional long-range quantum network is to demonstrate entanglement between a quantum processing node and a quantum repeater node. Since such nodes are often based on different hardware platforms, this task requires bridging the physical differences by designing compatible protocols and developing quantum interconnects \cite{awschalom2022roadmap}. Recent experiments have shown a connection between heterogeneous quantum systems using direct photon transmission \cite{maring2017photonic, Chai2026_hybridnetworklinking, Wang2026_heterogeneousentanglement}, while theoretical proposals have investigated ways to realize such a hybrid quantum link using photonic links and entanglement swapping \cite{sun2025hybrid, Tissot2025_hybridsingleion, Cussenot2025_unitingprocessingnodes}. Several demonstrations and proposals toward blind operations and RSP on remote quantum processors rely on direct photon transmission to the client \cite{vanDam2024,Drmota2024, Wei2025}, which can also be extended to multiple qubit RSP \cite{Propp2026}. 

\begin{figure}
    \centering
    \includegraphics[width=1.0\linewidth]{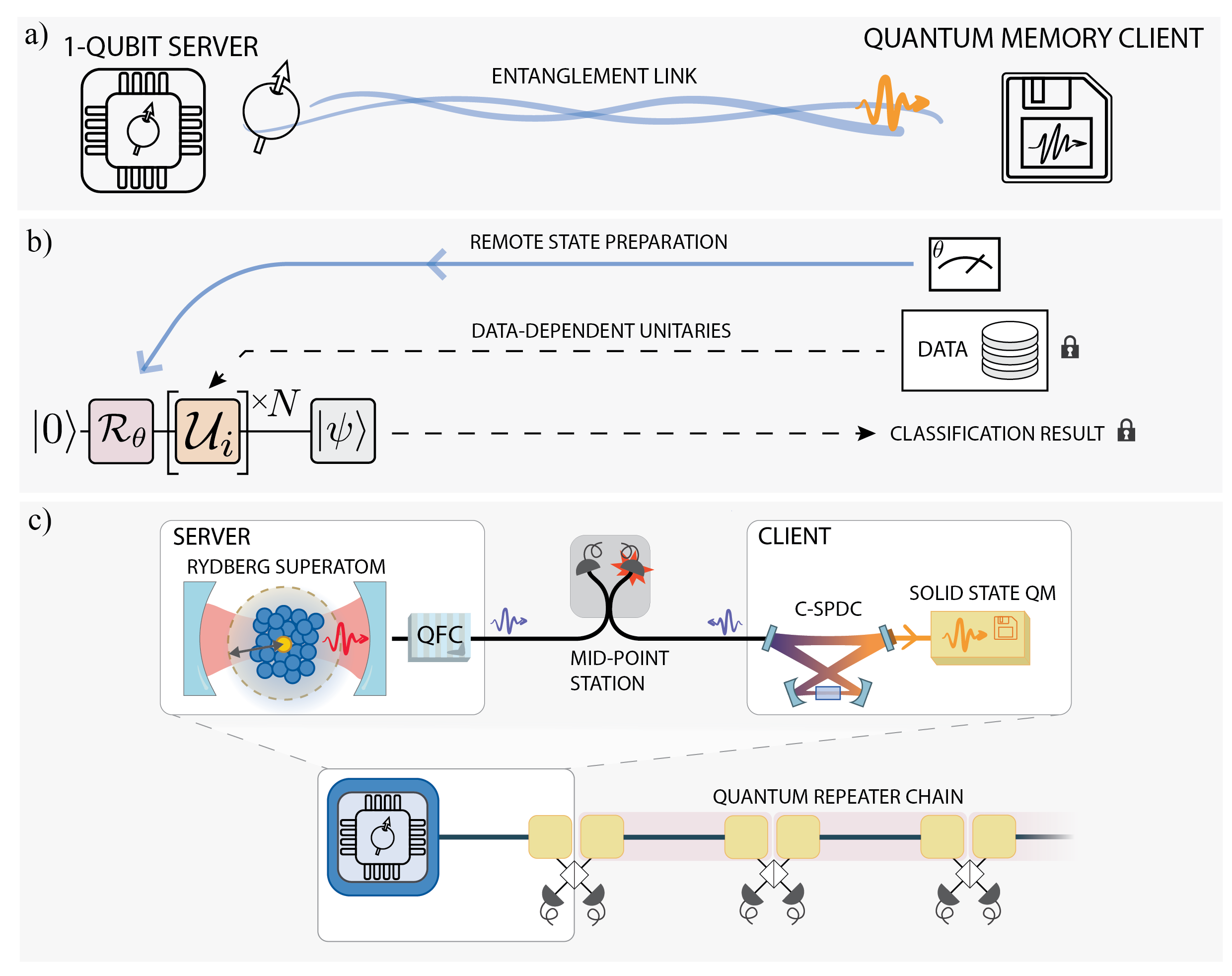}
    \caption{Schematic of the proposed application. a) We consider an elementary quantum network link between a server that has a single-qubit processor and a client consisting of a quantum memory. The two parties can share remote entanglement. b) A Blind Single Qubit Classifier protocol is implemented on the quantum server: the client first performs remote state preparation on the server's qubit, then they communicate classically a set of encrypted, data-dependent unitaries, which the server applies. The result of the classification, as well as the data, is unknown to the server. c) Proposed hardware implementation, using a Rydberg superatom as a one-qubit processor, and a rare-earth ion-based solid-state quantum memory (QM) at the client station. The entanglement is distributed between the server and client using a cavity-enhanced spontaneous parametric downconversion entanglement source (C-SPDC) and Bell-state measurement at a mid-point station with a photon generated by the server. This scheme is compatible with a quantum repeater architecture, enabling future long-distance operation of this application.}
    \label{fig:architecture_and_context}
\end{figure}

In summary, integration of blind, server-delegated, computing applications in a scalable,long-distance quantum network requires protocols and hardware that are functional for repeater-compatible entanglement generation and distribution. This work proposes a prototypical network capable of delivering an example of QML that is partially-blind, remote, and compatible with near-term quantum network hardware. Fig. \ref{fig:architecture_and_context} depicts the proposed network from its high- to low-level architectural components, as well as its compatibility with a quantum repeater chain. The proposal involves: (i) developing a networked protocol based on the single-qubit classifier \cite{PrezSalinas2020}, (ii) proposing a hybrid, repeater-compatible, quantum network link where the protocol developed in (i) could be implemented, and (iii) simulating and benchmarking a case-study for such protocol, namely the classification of credit card transaction fraud, using a real-life dataset. We also evaluate the long-term scenario where a multi-qubit classifier can be executed in a network, leading to stronger cryptographic properties and, potentially, classical simulation hardness. By introducing operational aspects of the network in this panorama, we show how the 2-qubit case already offers significant advantages such as full blindness and verifiability.

The paper is divided as follows. In Section II, the partially-blind SQC (PB-SQC) is introduced along with a quick review of the mechanisms that enable blindness in a quantum network. An extension of the PB-SQC protocol, the blind 2-qubit classifier, is also introduced and evaluated in the network context. Section III presents the short-term network that could execute PB-SQC, including the server, client, and intermediate node hardware infrastructure. Section IV presents the results of real classification data executed using a quantum network simulation environment. The inclusion of decoherence mechanisms in view of the analysis of Section III allows to establish achievable performances of such a network. Physical layer parameters such as entanglement generation rate, fidelity, and server-client distance, as well as high-level parameters such as time-to-service and classification metrics, are evaluated and discussed. Section V concludes the article discussing perspectives on the deployment of such proto-quantum-networks and their impact on the current quantum communication ecosystem worldwide.

\section{Partially-Blind Single-Qubit Classification Protocol}

\subsection{Blind Quantum Computing Preliminaries}

Blind quantum computing (BQC) \cite{broadbent2009universal, arrighi2006blind} is a potential application running over a quantum network whereby a \textit{client} may delegate the execution of specific tasks to a \textit{server} while maintaining privacy over the input data, measured outcome, and executed calculation. Central to the application are three functionalities: delegation, meaning the client can offload the calculation to the server; verifiability, which allows the client to verify that the server executed the specified task and no other; and noise robustness, the client's capacity to extract useful information from the outcome even in presence of a certain level of noise. As in Quantum Key Distribution (QKD)\cite{gisin2002quantum}, noise and maliciousness of a third party (including the server itself) are treated equally; in other words, a practical error threshold must be stipulated to guarantee robustness \cite{Gheorghiu2018}. In measurement-based blind quantum computation \cite{Broadbent2009}, a client may perform a private computation on the server such that the server does not learn anything about the computation besides an upper bound on its size. In the original framework, this is achieved by preparing many phase-rotated qubits on a server in a graph state (or alternatively, rotating many qubits held by the server in a blind fashion) and performing a measurement-based blind quantum computation.

In this work, focus is given to versions of blind quantum computing that can be put in practice on upcoming quantum networks of second generation (untrusted nodes equipped with quantum memories for light) \cite{wehner2018quantum}: blind data classification. In data classification, each element of a training dataset $d_T \in D_T$ has a known label $L(d_T)$ and a neural network is trained so that, with sufficiently high probability, the classification it outputs for the dataset matches the pre-known label $k \rightarrow L(d_T)$, at which point the neural network can be used to classify new data with the same features as the training data. A single qubit can represent a depth-one neural network with arbitrarily high dimensional input at the cost of an increasing circuit depth, removing any speed-based quantum advantage by making use of data re-uploading. These single-qubit classifiers are straightforward to implement, including in photonics \cite{Ding2023}, and can be iteratively built up to multi-qubit classifiers \cite{GarciaEscartin2013}. However, single-qubit classifiers are also interesting in their own right as testbeds for modified classification protocols to deliver, e.g., private classification, providing a quantum advantage in the form of privacy. With a single remote qubit on a server, it is possible to verifiably perform quantum classification with limited privacy; the server will know that a classification is taking place but not have access to the datasets being classified $d\in D$ nor the set of outcomes for each data element $C(d)$. We henceforth refer to it as partially-blind. Like blind quantum computation, this is enabled by arbitrary single qubit remote state preparation.

Here, we look at a form of blind quantum computing where the phase rotations are achieved via remote state preparation \cite{Bennett2001}, whereby a client may securely prepare an arbitrary single-qubit state on a server: $\ket{0} \rightarrow \tilde{U} \ket{0}$, with the unitary $\tilde{U}$ unknown to the server and chosen from a random set by the client. Typically, these random unitaries $\tilde{U}$ form a one-time pad so that a classically communicated rotation is sufficient to prepare an arbitrary single qubit state securely. In traditional measurement-based blind quantum computation, the remotely prepared states are chosen from two sets---those suitable for computation and those suitable for verification \cite{leichtle2021verifying}---though recent advances unify these into a single set \cite{Kapourniotis2024}. If the remote state preparation is secure, an untrusted quantum server cannot \emph{a priori} know which rounds are computations and which are verification, giving rise to verifiably private quantum computation.

\subsection{Quantum Classification}

\begin{figure}
    \centering
    \includegraphics[width=0.9\linewidth]{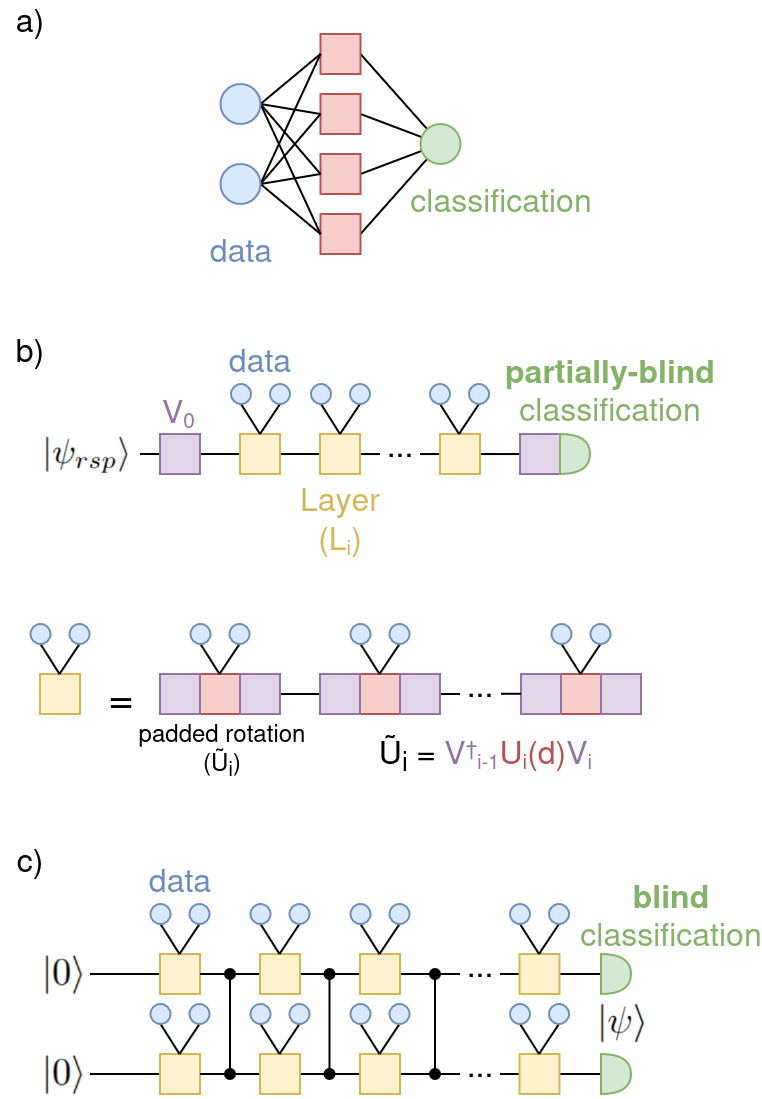}
    \caption{a) Neurons (squares) of a neural network are trained according to a known dataset so that new instances can be classified. b) A single qubit classifier uses data-re-uploading to adjust the rotation of qubit transformations so that, after data and training rotations, of the final state yields the classification result. Padding the data and training rotations with Haar-random unitaries ensures that a server cannot assess the user's data when transmitted over a classical channel, c) A blind two-qubit classifier uses the same mechanism of padding as in (b), with the difference that a two-qubit gate is applied between each unitary layer. As discussed in the manuscript text, blindness is ensured because the user can adjust the sequence of unitaries and implement any two-qubit circuit.}
    \label{fig:BSQC}
\end{figure}

In quantum classification (Fig. \ref{fig:BSQC}), the classification task is performed by a quantum computer. Here, a hybrid quantum-classical neural network is trained to map elements of a training data set $d_T \in D_T$ to unitaries $U(d_T)$ in such a way that, if a POVM $\{\hat{\Pi}_k\}$ is measured for a fiducial state transformed by the unitary $U(d_T)\ket{0}$, the outcome $k$ corresponding to the label $L(d_T)$ occurs with sufficiently high probability: ${\rm Tr} [U(d_T) \ket{0}\bra{0}U^\dagger(d_T)\hat{\Pi}_{L(d_T)}] \rightarrow 1 ]$.\footnote{Note that this is a slight oversimplification; classification could also be based on the distribution of measurement outcomes and not the outcomes themselves. However, for our purposes it suffices to consider the asymptotically deterministic case as it is the simplest to make blind.} Notably, in \cite{PrezSalinas2020} it was discovered that data-reuploading allows one to perform this quantum classification universally with fewer qubits and (remarkably) even a single qubit; the single unitary $U(d_T)$ is, instead, replaced by a sequence of unitaries $\prod_1^N U_i(d_T)$, with $N$ the depth of the classifier. Once trained, a classifier can be used to classify new data provided that data has the same features as the training data. 

Mathematically, we can summarize the implementation of a depth-$N$ single qubit classifier as follows: \begin{enumerate}
    \item In the training stage, a hybrid quantum-classical neural network is trained to find unitary-outputting functions $f_i: d\rightarrow U_i(d)$ such that, for each $d_T \in D_T$, ${\rm Tr} [(\prod_{i=1}^N U_i(d_T))\ket{0}\bra{0}(\prod_{i=1}^N U_i^{\dagger}(d_T))\hat{\Pi}_{L(d_T)}]>1-\epsilon$, with $\epsilon$ some allowable error. 
    \item In the classification stage, the same quantum-classical neural network is used to classify unknown data elements $d\in D$, where the probability of a classification is calculated by the Born rule $P_k ={\rm Tr} [(\prod_{i=1}^N U_i(d)) \ket{0}\bra{0}(\prod_{i=1}^N U_i^\dagger(d)) \hat{\Pi}_{k}]$.\\
    \label{item:a_priori}
\end{enumerate}

\subsection{Partially-Blind Single Qubit Classification}
\label{sec:sqc_theory}

Inserting blindness into the above protocol is straightforward using remote state preparation (RSP), which is used as a one-time pad that we propagate through the algorithm. We consider only blind classification without training. \footnote{For an analysis of blind-training using a measurement-based setting to implement gradient descent for BQC, see, e.g., \cite{pappa2025}.} This is enabled by using random bi-unitary conjugation of the data-dependent unitaries $U_i(d) \rightarrow \tilde{U}_i(d) = \tilde{V}_i U_i(d) \tilde{V}_{i-1}^\dagger$ for $0<i<n$; here, the $V_i$ are (Haar) randomly chosen padding unitaries used to hide our data-dependent unitaries $U_i(d)$. For the final unitary $U_n(d)$, we instead use a random equatorial rotation followed by a bit-flip for the left conjugation: $U_n(d) \rightarrow \tilde{U}_n(d) = \sigma_x^pR_Z\left(\frac{b\pi}{4}\right) U_n(d) \tilde{V}_{n-1}^\dagger$, $p\in\{0,1\}$ and $b\in[8]$. This ensures that the final measurement outcome (performed on the computational basis) leaves the server with no information about the protocol. 

By also remotely preparing a random equatorial state $\tilde{\ket{0}} = \tilde{V}_0\ket{0}:=R_Z\left(\frac{b'\pi}{4}\right) \mathbf{H}\ket{0}$, for $b'\in[8]$ and $\mathbf{H}$ the Hadamard gate \(\mathbf{H} = \frac{1}{\sqrt{2}}\begin{pmatrix} 1 & 1 \\ 1 & -1 \end{pmatrix}\), we find that 

\begin{widetext}\bea\label{correctness}
\tilde{P}_k &=& {\rm Tr} \left[\left(\prod_{i=1}^n\tilde{U}_i(d)\right) \tilde{\ket{0}}\tilde{\bra{0}}\left(\prod_{i=1}^n\tilde{U}_i^{\dagger}(d)\right)\tilde{\hat{\Pi}}_{k}\right]\nonumber \\
&=&{\rm Tr} \left[\sigma_x^pR_Z\left(\frac{b\pi}{4}\right) U_n(d) \tilde{V}_{n-1}^\dagger\prod_{i=1}^{n-1}\left(\tilde{V}_i U_i(d) \tilde{V}_{i-1}^\dagger\right)\tilde{V}_0\ket{0}\bra{0}\tilde{V}_0^\dagger\prod_{i=1}^{n-1}\left(\tilde{V}_{i-1} U_i^\dagger(d) \tilde{V}_{i}^\dagger\right) \tilde{V}_{n-1}
U_n^\dagger(d)  \sigma_x^p R_Z\left(\frac{b\pi}{4}\right)
\hat{\Pi}_{k}\right] \nonumber \\ 
&=&{\rm Tr} \left[\left(\prod_{i=1}^n U_i(d)\right) \ket{0}\bra{0}\left(\prod_{i=1}^n U_i^{\dagger}(d)\right)\sigma_x^pR_Z\left(\frac{b\pi}{4}\right)\hat{\Pi}_{k}R_Z\left(\frac{b\pi}{4}\right)\sigma_x^p\right], \quad\text{for}\: k\in\Im(L).
\eea\end{widetext}

Here, $\Im(L)$ is the set of possible classifications of the data. For $p=0,b=0$, this agrees with our expression in item \ref{item:a_priori} of II-B, i.e., the depth-N SQC, and for $p=1,b\in\{0,1,\dots7\}$, this swaps and rotates the outcomes for measurements made in the computational basis, ensuring correctness for the client with respect to the non-blind protocol. 

Blindness is ensured by the useful group property $\int_G \mathrm{d}V V = 0$, with $\mathrm{d}V$ the Haar measure of a group $G$. Because of this, the server learns nothing about the actual unitaries chosen, that is: in each step of the protocol the server learns $\tilde{V}_i U_i(d)\tilde{V}^\dagger_{i-1}$ but integrating over $\tilde{V}_i$ and $\tilde{V}^\dagger_{i-1}$ we see that this gives no information about the underlying unitary $U_i(d)$ because $\int_G \mathrm{d} V_i\mathrm{d} V_j^\dagger V_i U(d)_i V_j^\dagger = 0$ for $i\neq j$. The same holds for the last unitary $\int_G \mathrm{d} V^\dagger \sigma_x^pR_Z\left(\frac{b\pi}{4}\right)U_n(d)V^\dagger =0$.

If the server chooses to deviate from the protocol and attempts to learn any unitary $\hat{U}_j:=U_j\cdots U_1(d)$ for $j\in[n]$ by making measurements on the server state, the unitary twirling theorem $\left(\int_G \mathrm{d}V V \rho V^\dagger = {\mathrm Tr}(\rho) \id / 2\right)$ ensures that they will learn nothing data-dependent:
\begin{equation}
    \begin{split}
    &\int_G \mathrm{d}V V \hat{U}_jR_Z\left(\frac{b'\pi}{4}\right)\mathbf{H}\ketbra{0}\mathbf{H}R_Z\left(\frac{b'\pi}{4}\right)\hat{U}_j V^\dagger \\
    &= \Tr[\hat{U}_jR_Z\left(\frac{b'\pi}{4}\right)\mathbf{H}\ketbra{0}\mathbf{H}R_Z\left(\frac{b'\pi}{4}\right)\hat{U}_j]\frac{\id}{2}\\
    &=\frac{\id}{2},
    \end{split}
\end{equation}
because unitary evolution preserves the trace of quantum states.

Finally, we see that $\sum_{p,b} \expect{0/1 | \sigma_x^pR_Z\left(\frac{b\pi}{4}\right) U(d) |0}=0$ i.e.\ the measurement statistics yields no information to the server in the absence of knowledge of the padding unitaries applied. We also see that $\int_G \mathrm{d}V_i \mathrm{d}V_j^\dagger V_iUV_j^\dagger=0$ and $\int_G \mathrm{d}V_j^\dagger \sigma_x^p U(d)V_j^\dagger=0$, i.e., the classical communication of padded unitaries reveals no information about the unpadded unitaries and thus also no information about the data $d$. If the RSP is done securely, the server has no way of uncovering the information about the (unpadded) unitaries performed and exactly no information is leaked. Note that while we use Haar random padding unitaries $V_i$ for ease of proof, it is sufficient to replace these with random bitflips $\sigma_x^{p_i}$ and $z$-axis rotations $R_Z(\frac{b_i \pi}{4})$, as these form a $1$-design \cite{Gross2007}. The complete structure of the Partially-Blind Single-Qubit Classifier (PB-SQC) is presented as a pseudo-code in Alg. \ref{alg:PB-SQC}.

\begin{algorithm}[H]
\caption{Partially-Blind Remote Noise-robust Single-Qubit Classification}
\label{alg:PB-SQC}
    \begin{algorithmic}
    \Require $U^{model} \gets$ \textsc{classical training}
    \Require $U^{data} \gets$ \textsc{new unclassified instance}
    \State $M \gets$ \textsc{number of rounds}
    \State $n \gets$ \textsc{data reuploading steps}
    \State $m \gets 1$
    \State $B \gets$ \textsc{M-long random binary vector}
    \State $b \gets$ \textsc{M-long random vector with entries 0...7}
    \While{$m < M$}
        \State
        \State \textbf{Run Proto Quantum Network}
        \While{\textbf{not}(Client$\leftrightarrow$Server Entangled)}
            \State Client$\leftrightsquigarrow$Server: Entanglement distribution 
        \EndWhile
        \State
        \State \textbf{Run PB-SQC Instance}
        \State Client $\rightsquigarrow$ Server: Arbitrary-basis RSP preparing $V_0 \ket{0}$
        \State Client: choose $n$ Haar-Random Unitaries $\{V\}_{i=1..n}$
        \For{$i = 1..n-1$}
            \State Client $\Rightarrow$ Server: $V_i U^{data}_i U^{model}_i V_{i-1}^{\dagger}$
            \State Server: executes rotation
        \EndFor
        \State Client: $\theta = \frac{\pi b[m]}{4}$
        \State Client $\Rightarrow$ Server: $\sigma_x^{B[m]} R_Z\left(\theta\right) U^{data}_n U^{model}_n V_{n-1}^{\dagger}$
        \State Server: executes rotation and measures
        \State Server $\Rightarrow$ Client: measurement outcome
        \State
        \State $m=m+1$
    \EndWhile
    \State Client only accepts if at least half of the outcomes coincide
    \end{algorithmic}
\end{algorithm}

We will now elaborate on the noise robustness of the PB-SQC. Let us assume that the two-qubit channel fidelity is $F$, i.e., the fidelity associated with the distributed entangled state over the network between Server and Client. After RSP, the state on the Server can be model this as a perfectly-initialized state experiencing the effect of a depolarizing channel with error $\zeta_S:=\tfrac{4}{3}(1-F)$; here, the factor $\tfrac{4}{3}$ is due to the fact that the channel acts on the two-qubit state and the subscript $_S$ refers to the \textbf{s}ingle qubit on the Server. Then, the initial state on the Server's register takes the form $(1-\zeta_S)\tilde{V}_0\ketbra{0}\tilde{V}_0^\dagger+\zeta_S\id_2$. Moreover, assuming a depolarizing noise model where the execution of each unitary is modeled as the application of an independent and identically distributed depolarizing channel of error $\delta$, by linearity of quantum mechanics, the probabilities of each measurement outcome $k\in \Im(L)$ of the Server, on each of the $M$ rounds, will be
\begin{equation}
\begin{split}
    \Pr(k)= &(1-\delta)^n(1-\zeta_S)\tilde{P}_k\\
          &+(1-(1-\delta)^n(1-\zeta_S))\mathrm{Tr}[\id_2\hat{\Pi}_k];
\end{split}
\end{equation}
this does not give the server any additional information in comparison to the noiseless case. Given that the classification outcome of data $d$ is $k=L(d)$, consider the identical Bernoulli variables $X_1,\ldots,X_M$ denoting if a round outcome has been successful, i.e.\ in round $i\in[M]$, $X_i=1$ iff the outcome is $k=L(d)$. Then the sum of these variables $\mathcal{X}=X_1+\cdots+X_M$ is a Binomial variable counting the number of rounds where the successful outcome was obtained, $\mathcal{X}\sim\mathrm{Bin}(M,\mathcal{P}_{L(d)})$. By Hoeffding's inequality, if $1/2\leq \mathcal{P}_{L(d)}$, the probability that the protocol succeeds is exponentially increasing in the number of rounds
\begin{equation}
    \begin{split}
        &\Pr(\mathcal{X}\geq M/2) \\
    &= \sum_{m > M/2}^M\binom{M}{m}(\mathcal{P}_{L(d)})^m(1-\mathcal{P}_{L(d)})^{M-m}\\
    &\geq 1-\exp(-2M\left(\mathcal{P}_{L(d)}-1/2\right)^2)\\
    &\geq 1-\exp(-2M\left((1-\delta)^n(1-\zeta)\tilde{P}_{L(d)}-1/2\right)^2),
    \end{split}
    \label{eq:prob_succ_eq}
\end{equation}
which ensures noise robustness provided the client can execute the protocol a sufficiently large number of times.

\subsection{Verifiable Blind Two Qubit Classification}
\label{sec:tqc}

One of the hallmarks of BQC is the ability to go beyond blindness and verify that the server produces sensible results. This, along with the ability to diagnose the noise levels in the remote server, relies on trap (or test) rounds where the client can easily predict the outcomes of tests \cite{broadbent2009universal}. To add verifiability and noise robustness, we propose scaling up to a blind two-qubit classifier (B-TQC).
  
The necessity of two qubits is due to the two-colorability of the two-qubit graph, so that we can use the trap schemes of standard BQC \cite{broadbent2009universal}. Here, either two single-qubit RSPs or a single multi-qubit RSP \cite{Propp2026} are used to provide the one time pads necessary to hide an arbitrary two-qubit gate, which has a simple KAK decomposition 
\begin{multline}\label{twoqubit}
    U_i(d) = (U_{1,i}(d) \otimes U_{2,i}(d))\,\mathbf{CZ}\,(U_{3,i}(d) \otimes U_{4,i}(d)),\\
    U_i \in U(2),
\end{multline}
where the $U_{k,i}(d)$ are four data-dependent single-qubit unitaries acting on top ($k=1,3$) or bottom ($k=2,4$) qubits, and $\mathbf{CZ}$ is the standard two-qubit $CZ$ gate. We use the same padding scheme as the single qubit gates: $U_i(d) \rightarrow \tilde{U}_i(d) = \tilde{V}_i U_i(d) \tilde{V}_{i-1}^\dagger$ for $i<n$ and $U_n(d) \rightarrow \tilde{U}_n(d) = \left(\sigma_{x,1}^{p_1} R_{Z,1}(\frac{b_1\pi}{4}) \otimes\sigma_{x,2}^{p_2} R_{Z,2}(\frac{b_2\pi}{4})\right)U_n(d) \tilde{V}_{n-1}^\dagger$ for the final unitary, with the different that, now, $\{V\}_{i}$ are Haar-Random two-qubit unitaries (2QUs). Here as in the single qubit case, for the final step the client uses two bit-flips and Z-axis rotations on each qubit to pad the final unitary. Also, as in the single qubit case, the Haar-random unitaries can be substituted for a suitable $1$-design. 

\begin{algorithm}[H]
\caption{Verifiable Noise-Robust Remote Two-Qubit Classification}\label{alg:1}
\begin{algorithmic}
\Require $U^{model}$: classical training
\Require $U^{data}$: new unclassified instance
\State
\State \textbf{\underline{Measurement phase:}}
\State $c \gets$ \textsc{number of computation rounds}
\State $t \gets$ \textsc{number of test rounds}
\State $B_1 \gets$ \textsc{M-long random binary vector}
\State $B_2 \gets$ \textsc{M-long random binary vector}
\State $b_1 \gets$ \textsc{M-long random vector with entries 0...7}
\State $b_2 \gets$ \textsc{M-long random vector with entries 0...7}
\State Client chooses a random partition $(C,T)$ of $[M]$ 
\For{$m \in [c+t]$}
    \State
    \State \textbf{Run Proto Quantum Network}
    \While{\textbf{not}(Client$\leftrightarrow$Server Entangled)}
        \State Client$\leftrightsquigarrow$Server: Entanglement distribution 
    \EndWhile
    \State
    \State \textbf{Run PB-SQC Instance}
    \State Client: Choose $n$ Haar-Random 2QUs $\{V\}_{i=0\ldots n-1}$
    \State Client $\rightsquigarrow$ Server: Arbitrary-basis RSP preparing $V_0\ket{00}$
    \If{$m\in C$ (computation)}
        \For{$i = 1\ldots n-1$}
            \State Client $\Rightarrow$ Server: $V_i U^{data}_iU^{model}_i V_{i-1}^{\dagger}$
            \State Server: executes rotation
        \EndFor
        \State Client: $\theta = \frac{\pi b_1[m]}{4}$
        \State Client: $U_{XZ} = \sigma_{x,1}^{B_1[m]} R_{Z,1}\left(\theta\right)  \sigma_{x,2}^{B_2[m]} R_{Z,2}\left(\theta\right)$
        \State Client: $U_{2\mathrm{QU}} = U_{XZ}U^{data}_n U^{model}_n V_{n-1}^{\dagger}$
        \State Client $\Rightarrow$ Server: $U_{2\mathrm{QU}}$
        \State Server: executes rotation and measures
    \EndIf
    \If{$m\in T$ (test)}
        \For{$i = 1\ldots n-1$}
            \State Client $\Rightarrow$ Server: $V_iV_{i-1}^{\dagger}$
            \State Server: executes rotation
        \EndFor
        \State Client $\Rightarrow$ Server: $\sigma_{x,1}^{B_1[m]}\sigma_{x,2}^{B_2[m]}V_{n-1}^{\dagger}$
        \State Client $\Rightarrow$ Server: $\{\sigma_{x,1}^{B_1[m]}\sigma_{x,2}^{B_2[m]} \hat{\Pi}_k \sigma_{x,1}^{B_1[m]}\sigma_{x,2}^{B_2[m]}\}$
        \State Server: executes rotation and measures
    \EndIf
    \State
    \State Server $\Rightarrow$ Client: measurement outcomes
\EndFor
\State
\State \textbf{\underline{Post-processing phase:}}
\State $\tau \gets$ \textsc{acceptance threshold}
\State $t_{\rm fail} \gets$ \textsc{number of failed test rounds}
\State $\Omega \gets$ \textsc{number of outcomes}
\State $\omega \gets$ \textsc{number of coinciding outcomes}
\If{$t_{\rm fail}\geq \tau$}
    \State Client: aborts protocol
\Else
    \If{$\omega \leq \tfrac{\Omega}{2}$}
        \State Client: accepts
    \Else
        \State Client: rejects
    \EndIf
\EndIf
\end{algorithmic}
\end{algorithm}

Now, we can additionally run trap rounds, where one of the qubits is decoupled from the other by being prepared (and kept) in a computational basis state (dummy state). This yields the same security checks as in \cite{broadbent2009universal}, and can also be extended to more qubits in the same manner. Noise robustness is achieved, as in the PB-SQC, by repeating the protocol many times and using majority voting. The client is now able to verify the reported noise by the server, i.e., delegated benchmarking \cite{Chen2024}.

Unlike its single qubit counterpart, the B-TQC achieves full blindness because the sequence of unitaries transmitted could implement any two-qubit circuit. Furthermore, we only require two RSP qubit states to implement it, independently of circuit depth $N$, the $N$ unitaries comprising the classification. In comparison, a fully measurement-based implementation of a TQC requires $\mathcal{O}(N)$ RSP qubits \cite{Fleur2025BQML}. Instead, our B-TQC is constrained by memory coherence time during which the classification is performed, as well as (potentially) by composability of security as a subroutine \cite{Broadbent2009, Dunjko2014, Maurer2011, Kashefi2017}. In this work, we focus on the hardware implementation of the PB-SQC since it is compatible with already mature and demonstrated hardware platforms.

Finally, we end this section presenting Table \ref{table:requirements}, which associates application functionalities and different protocols discussed here in order of increasing complexity and generality. Note that in going to multiple qubit classifications, the remote preparation of qubits can either be done individually or collectively as in Ref. \cite{Propp2026}, utilizing the qudit structure of Rydberg superatoms as in Ref. \cite{Burshtein2026}.

\begin{table}[h]
    \centering
    \caption{Protocol Classification w.r.t. Application Functionalities} \label{table:requirements}
    \setlength\tabcolsep{5pt}
    \footnotesize\centering
    \begin{tabular}{c|c|c|c|c}
    Protocol & Blind & Noise Robust & Verifiable & Delegated\\
    \hline
    \hline
      PB-SQC & Partially & Yes & No  & Yes\\
      B-TQC & Yes & Yes & Yes & Yes\\
    B-QML   & Yes       & Yes & Yes & Yes\\
      BQC    & Yes       & Yes & Yes & Yes\\
    \hline
    \hline
    \end{tabular}
    \vspace{-4pt}
    \begin{flushleft}
    \scriptsize{$^{\ast}$ PB-SQC: Partially-Blind Single-Qubit Classifier, B-TQC: Blind Two-Qubit Classifier, B-QML: Blind Quantum Machine Learning, BQC: (Universal) Blind Quantum Computation.}
    \end{flushleft}
\end{table}

\section{Quantum Network Hardware for Repeater-compatible PB-SQC}

\begin{figure}[h]
    \centering
    \includegraphics[width=\linewidth]{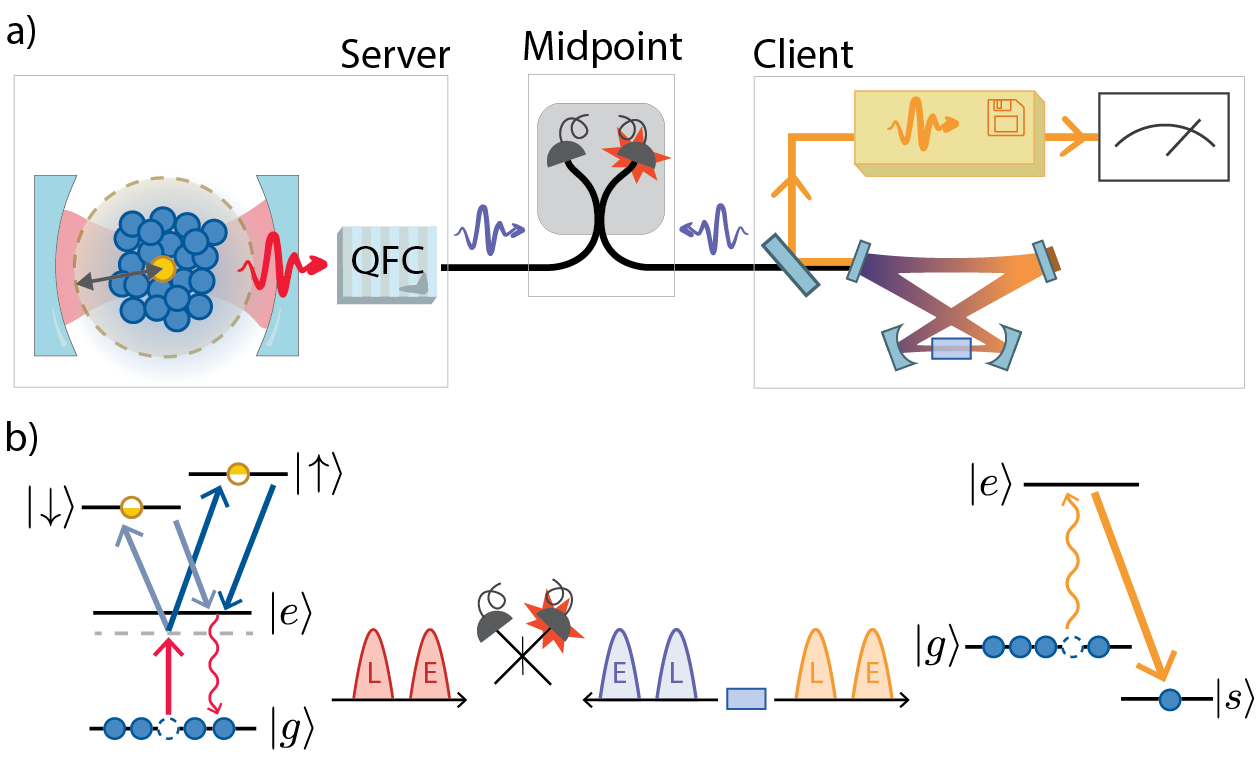}
    \caption{a) Schematic of the proposed network link hardware. The server consists of a Rydberg superatom qubit embedded in a cavity, and a quantum frequency conversion stage to convert the output photon to the telecom band. The client node consists of an entanglement source that generates a photon pair. One photon, at telecom wavelength, is sent to the midpoint station, the second is stored in an absorptive quantum memory. The midpoint station performs a Bell state measurement which leads to entanglement swapping. b) Simplified level scheme of the Rydberg superatom and the quantum memory, and encoding of the photons.}
    \label{fig:netArch}
\end{figure}

In this section, we introduce a blueprint for a near-term quantum network link that supports the operation of the proposed PB-SQC scheme. A crucial difference between our proposed implementation and other experimental demonstrations of RSP is that we use entanglement swapping rather than direct photon transmission to distribute entanglement between the server and client \cite{Wei2025, Drmota2024}. This makes the architecture compatible with integration into a quantum repeater chain, enabling long-distance operation in the future \cite{Cussenot2025_unitingprocessingnodes}.

\subsection{Hardware Description}\label{sec:hardware}

The hardware link can be divided into three main elements: the \textit{server}, the \textit{client}, and a \textit{mid-point station}.\\

\textit{Server -} For this prototype application, the server has two basic requirements: single-qubit rotations and (deterministic and efficient) light-matter entanglement generation. A Rydberg superatom \cite{kumlin2023_quantumoptics, Ebert2015_coherence} combines precise control of a qubit encoded in the collective state of a small atomic ensemble \cite{Spong2021_collectivelyencoded} with collectively-enhanced light-matter entanglement generation \cite{An2025entangling}. For these reasons, it is the Server's platform of choice in the proposed architecture. We consider the protocol used in \cite{Sun2022deterministic} to generate entanglement between a superposition of collective atomic states and time-bin degrees of freedom of a photon. Two-laser transitions (red and blue upwards arrows in Fig. \ref{fig:netArch}b) couple the ground state to two Rydberg states ($\ket{g}$ and $\ket{\uparrow/\downarrow}$, respectively) and are used to prepare the ensemble in a superposition. First, a de-excitation control pulse (dark blue arrow pointing downwards) is sent resonantly to the $\ket{\uparrow}\to\ket{e}$ transition, causing the conditional emission of a photon on time-bin "early". The cloud is prepared again in $\ket{\uparrow}$ to preserve the superposition state. Finally, a second pulse (light blue downwards-pointing arrow) conditionally de-excites the atoms from state $\ket{\downarrow}$, resulting in a photon in the "late" time-bin. After preparing again the atoms in $\ket{\downarrow}$, this results in the following light-matter entangled state:
\begin{align}
\begin{split}
    |\psi_\textrm{server}\rangle  &= \frac{1}{\sqrt{2}}(\hads{r,e}|\uparrow\rangle + \hads{r,\ell}|\downarrow\rangle) \\
    &= \frac{1}{\sqrt{2}} (|\uparrow e\rangle + |\downarrow \ell \rangle),
\end{split}
\label{eq:psi_server}
\end{align}
where $\uparrow(\downarrow)$ represents the superatom qubit state, and $e\left(\ell\right)$ is the Early (Late) time-bin mode of the photon. The coherence time of the Rydberg-encoded qubit is limited to a few tens of microseconds due to the high sensitivity of these states to external electric noise, the short relaxation lifetime, and motional dephasing. Nevertheless, after light-matter entanglement generation, the single collective excitation can be transferred to a long-lived Zeeman sub-level of the atom's ground state \cite{Han2010_quantumrepeatersrydberg}, allowing for extended qubit coherence times. Manipulating the superatom in order to involve a larger number of states addressable via optical or microwave signals can, in principle, enable the preparation of higher-dimensional states, or qu\textbf{d}its \cite{Jiao2025_photonicqutrit}. This further positions the Rydberg superatom as a long-term candidate for the implementation of the B-TQC protocol introduced in Sec. \ref{sec:tqc}. Finally, we introduce a quantum frequency conversion (QFC) step to convert the server photon wavelength, emitted at the Rubidium atomic transition frequency corresponding to 780nm \cite{Albrecht2014_waveguide}, to the telecom C-band for more efficient transmission through the fibre network.
\\

\textit{Client -} The client node should be able to establish entanglement between one of its photons and the Server's Rydberg-encoded qubit. This photon will then be measured in an arbitrary basis to perform remote state preparation (RSP) on the Server's qubit. Due to its potential for multiplexing over several degrees of freedom and, thus, high achievable entanglement distribution rates, the combination of a source of entangled photon-pairs based on cavity-enhanced spontaneous parametric downconversion (C-SPDC) and an absorptive quantum memory based on rare-earth ion ensembles in a solid-state host crystal (REI-QM) is the platform assigned to the Client. It is important to highlight that the Client node's hardware choice is naturally compatible with a multiplexed quantum repeater chain as depicted in Fig. \ref{fig:architecture_and_context}. The Client source produces the quantum state\cite{Hellebeck2024_multimodenature}:
\begin{equation}
    |SPDC\rangle = \bigotimes_{r=1}^R \sqrt{1-\xi^2} \sum_q^\infty\xi^q (a_{s,r}^\dagger a_{i,r}^\dagger)^q |vac\rangle,
\end{equation}
which is a tensor product of two-mode squeezed states for each photonic mode $r$. Here, $s(i)$ indicates the signal (idler) photon for each mode. The parameter $\xi$ is the brightness of the SPDC source, and can be tuned by manipulating the pump laser's power. Defining two time-bin modes and considering low brightness, the state of the SPDC source can be written as

\begin{align}
\begin{split}
    |\psi_{SPDC}\rangle = &(1-\xi^2) \big[ 1 + \xi (\hads{s,e} \hads{i,e} + \hads{s,\ell} \hads{i,\ell}) +\\
&+ \xi^2 (\hads{s,e} \hads{i,e})^2 + \xi^2 (\hads{s,\ell} \hads{i,\ell})^2 + \\
&+\xi^2 \hads{s,e}\hads{i,e}\hads{s,\ell}\hads{i,\ell} +O(\xi^3) \big] |vac\rangle.
\end{split}
\label{eq:psi_spdc}
\end{align}

The source is engineered to produce pairs of photons with two different wavelengths: the signal photon is compatible with the Pr$^{3+}$:YSO quantum memory (606nm \cite{Hanni2025_heraldedentanglement}), and the idler is in the telecom C-band to facilitate low-loss transmission over the fibre network. The cavity is used to ensure spectral matching of the photons with the bandwidth of the QM. The signal photon is stored as a collective excitation in the memory using a spin-wave atomic frequency comb (SW-AFC) protocol \cite{Hanni2025_heraldedentanglement, Minar2010_spinwavestorage, Rakonjac2021_entanglement}, effectively generating light-matter entanglement between the internal QM state and the idler photon, which to first order can be written as:
\begin{align}
    \ket{\psi_\textrm{SPDC-QM}} \approx \ket{vac} + \xi (\ket{\kappa_e e_i} + \ket{\kappa_\ell \ell_i}).
\end{align}
Here, we denote with $\kappa_{e\left(\ell\right)}$ a spin-wave stored in the QM (i.e. a collective excitation between states $\ket{g}$ and $\ket{s}$ in Fig. \ref{fig:netArch}), for a time mode $e(\ell)$. The subscript $i$ identifies the idler photon generated by the SPDC.
The AFC protocol implemented in REI-QMs \cite{afzelius2009multimode} is naturally compatible with time-multiplexing, which, besides making it a flexible platform for interfacing with different processing nodes, could also be used in the case of a qu\textbf{d}it where the server's light-matter entangled state would be encoded in multiple (\textbf{d}) time-bins as proposed for many-qubit RSP in Ref.\cite{Propp2026}. The multiplexing capability of the memory could also be used to speed up the entanglement generation in the case where multiple server qubits are used \cite{tissot2025_singledoubleclick}.
\\

\textit{Midpoint station -} A mid-point station equipped with a beam splitter and single photon detectors has the task of performing a Bell state measurement (BSM) on the joint photonic state of the server and client, which, upon heralding of a specific photon detection pattern, swaps the two light-matter entangled states to matter-matter entanglement \cite{stolk2024metropolitan, lago2021telecom, yu2020entanglement}. For example, detection of the singlet state ($\ket{\Psi^+}$, a pattern of two time-resolved detections on a single detector) yields the final Bell state: 
\begin{align}
\ket{\psi_\textrm{ent}} = \frac{1}{\sqrt{2}}(\ket{\uparrow \kappa_e} + \ket{\downarrow \kappa_{\ell}}).
\label{eq:psi_entangled}
\end{align}
For the BSM to be successful, the incoming photons from both nodes must be indistinguishable. While frequency indistinguishability is granted by the QFC, the temporal mode of the two photons can be matched by waveform shaping in the atomic emission and by time filtering.
\\

With this prototype network architecture, we aim to address three relevant aspects for future scalability to a long-distance, advanced quantum network: node heterogeneity; compatibility with repeater chains; and multiplexing. Future quantum networks will rely on diverse hardware platforms and, thus, applications designed to be compatible with a hybrid architecture will have a deployment advantage over homogeneous architectures. Our design includes frequency conversion \cite{Albrecht2014_waveguide} and a non-degenerate entangled photon source, which act as quantum interconnects between the server and client nodes, enabling control and engineering of the photonic states to achieve compatibility. Importantly, our protocol could be implemented on a server based on different hardware. In this design, a SPDC source paired with a quantum memory is used as a client node, but the same architecture can also be used as an intermediate node of a quantum repeater chain. The use of dual-rail encoding for the photonic qubit is compatible with a functional quantum repeater architecture \cite{Chou2007_functional}.

In summary, our protocol is compatible with a quantum link based on entanglement swapping, which is crucial for enabling multi-user PB-SQC, i.e., one Server connected to multiple Clients. This is further strengthened by the inclusion of Client hardware platforms that enable multiplexing in different degrees of freedom, ideal for high-rate automated quantum repeater chains \cite{askarani2021entanglement} and extensions making use of quantum multiplexing \cite{LoPiparo2019, piparo2020quantum, Xie2023, Propp2025}.

\subsection{Experimental Sequence and Performance of the Quantum Network Link}
\label{sec:network_performance}

\begin{figure}[h]
    \centering
            \includegraphics[width=\linewidth]{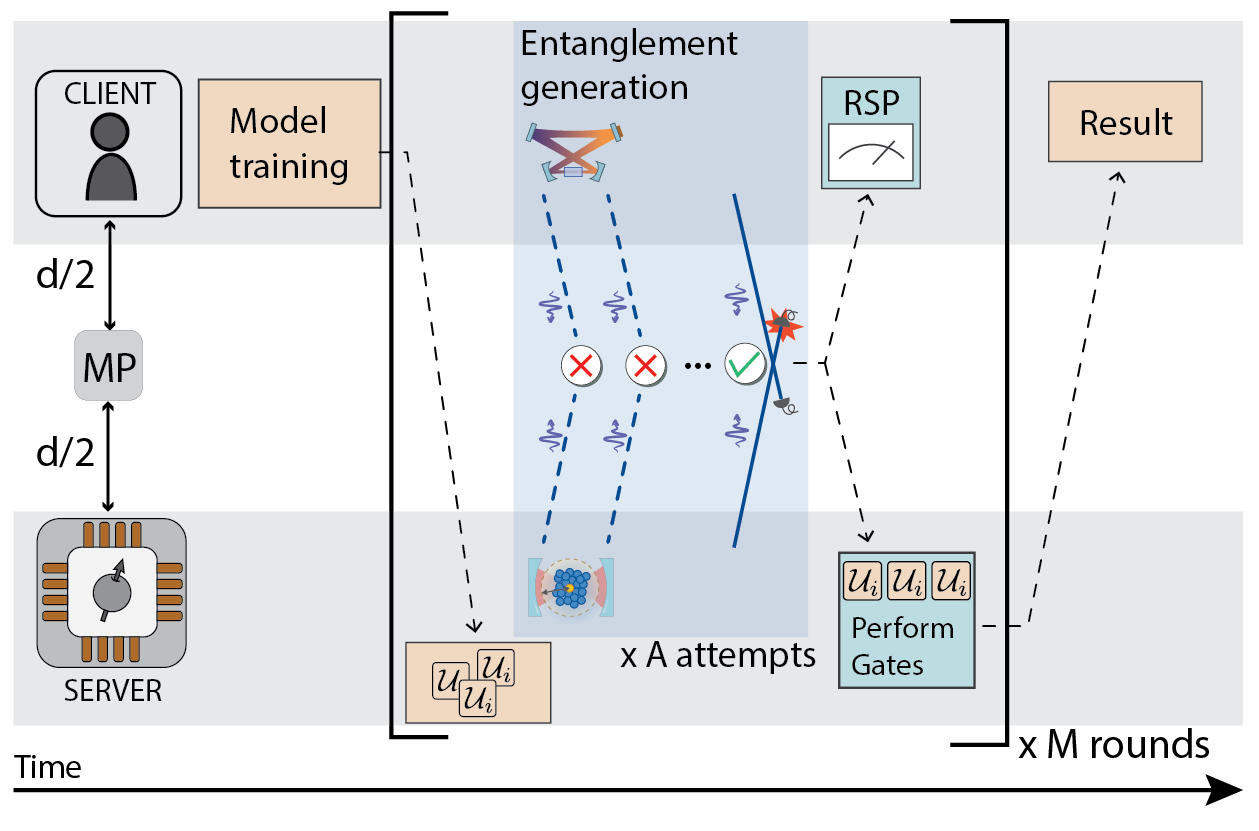}
            
    \caption{Sequence and timeline of the B-SQC protocol on a quantum network link. Before starting the protocol, the client perform the classical model training. The data-dependent encoded unitaries are sent to the server via a classical channel (black dashed arrow), while through the quantum channel (blue lines) the client and server attempt to generate entanglement. When entanglement is heralded, after $N$ attempts, the classical signal from the midpoint is sent to the two parties. At this stage, the client performs RSP on the server's qubit, and the server starts to perform the protocol. Finally, the server sends the classification result to the client and the protocol is repeated for the necessary amount of shots to guarantee the optimal performance..}
    \label{fig:expSeq}
\end{figure}

The proposed quantum network architecture is compatible with the PB-SQC protocol presented in Algorithm \ref{alg:PB-SQC}, which can be coordinated with network-level management of Client, Server, and Mid-Point Nodes. Fig. \ref{fig:expSeq} outlines how the PB-SQC protocol can be implemented in the envisioned quantum network. Before running the protocol, the Client trains the classical model in order to determine the single-qubit rotations $U_i\left(d\right)$. Training is executed once and the resulting model can be used to generate a set of data-dependent unitary operations for the Server to perform on every run of the protocol. These unitary operations are padded to ensure security, as explained in Sec. \ref{sec:sqc_theory}. When the protocol starts, the information on the padded set of unitaries is sent to the Server over a classical channel. At the same time, Client and Server initiate a sequence of entanglement generation attempts over the quantum channel. After a (variable) number of trials, $A$, the mid-point station will eventually herald a successful entanglement swapping round. The entanglement shared by the two parties is used by the Client to perform RSP on the Server's qubit by retrieving and measuring the photon stored in the QM on a randomly chosen basis, as introduced in Sec. \ref{sec:sqc_theory}. The Server now performs the gates received by the Client, measures the qubit in the computational basis, and sends the classification result to the Client. Depending on the level of noise, the Client runs the above steps $M$ times to ensure noise-robustness.

The entanglement generation stage is the most time-consuming during network operation and the fidelity of the shared Client-Server entanglement directly affects the RSP and, therefore, the number of repetitions $M$ of the protocol. In order to estimate how the protocol will perform on the proposed quantum network link, we develop a quantum optical model of our system and numerically calculate the success probability and fidelity. Starting from the quantum states defined in Eqs. \ref{eq:psi_server} and \ref{eq:psi_spdc}, we compute the outcome of the BSM operation and the state obtained by heralding on the expected detection pattern. We follow an approach similar to the one used in \cite{Hermans2023} to evaluate the effect of the multiphoton components generated by the SPDC source, loss on the photonic channels, and dark count noise on the detectors. The details of the modeling and calculations can be found in Appendix B. We note that there is a tradeoff between fidelity and success probability of the entanglement protocol as the distance changes, and that tuning the brightness of the SPDC source can lead to distance-specific optimized network operation.\\

Modelling the quantum network link hardware can provide insight into the expected performance of the PB-SQC application in a near-term, realistic setting. However, this will depend as well on the quality of the classically trained algorithm and the computation on the server's side. To complete this analysis, we introduce a case study where we combine classical model training on a real-life dataset and simulation of the server's operations.

\section{Case Study: Credit Card Fraud Detection with PB-SQC}
\label{sec:fraud_results}

Credit card fraud detection is a relevant problem for financial transaction operators \cite{neto2026fraud}. The object of this case study is the classification of real-world transactions into fraudulent and non-fraudulent -- see Appendix A, Subsection 1 for further details on the dataset. The performance evaluation of the PB-SQC takes three steps, described below, and further analyzed in subsections V-A (\textit{Numerical Results}) and V-B (\textit{PB-SQC Performance over a Prototype Hybrid Quantum Network}).
\begin{itemize}
    \item \textbf{Dataset Preparation}: From a 3054-instances-long pre-processed and balanced dataset (equal amounts of fraudulent and non-fraudulent transactions), two distinct datasets are created, henceforth referred to as \textsc{training} (2454-instances-long) and \textsc{testing} (600-instances-long) datasets.
    \item \textbf{Training}: Using a \textit{supervised learning} strategy \cite{kotsiantis2007supervised}, where the instance labels are provided, a classical optimizer adjusts the rotations with the goal of associating fraudulent instances to final states ($\ket{\psi}_d = \prod_{i=1}^N U_i\left(d_T\right)\ket{0}$) with high fidelity with respect to the state $\ket{1}$ and non-fraudulent to the state $\ket{0}$ -- refer to Appendix A, subsection 2, for further details.
    \item \textbf{Evaluation}: Using the protocol described in \textbf{Algorithm 1}, the classification results extracted using the \textsc{testing} dataset are interpreted using metrics derived from binary classification literature with particular focus on the so-called \textbf{f1-score} metric -- refer to Appendix A, subsection 1, for further details.
\end{itemize}

\subsection{Numerical Results}

In the particular case of binary classification and no gate-error, the final measurement performed by the server is a computational basis measurement with outcomes $k\in\{0,1\}$. In this case, there is no probability of aborting as the protocol will output the maximum outcome out of the two classes, according to Eq. \ref{eq:prob_succ_eq}. After one round, the probability associated with the measurement outcome of a given instance $d$ will depend on the channel decoherence. The two-qubit depolarizing channel with fidelity $F$ will introduce a single-qubit error $\zeta_S$ on the Server after RSP, as discussed in Section II-C, leading to:
\begin{equation}\label{eq:prob_depol}
    \textrm{P}_d(k) = (1-\zeta_S)|\bra{k}\ket{\psi}_i|^2+\frac{\zeta_S}{2},\quad\text{for}\:\:k\in\{0,1\}.
\end{equation}
While the above equation refers to a single measurement round, it is possible to calculate the probability of performing a correct classification for instance $d$ after $M$ rounds considering that the classification results are based on a majority vote decision, broken randomly for even-M, which is given by:
\begin{equation}\label{eq:P_afterM}
\begin{split}
    \textrm{P}_d(M,k)&:= \\
    \sum_{m=\tfrac{M}{2}+1}^M&\binom{M}{m}\left(P_d(k)\right)^m\left(1-P_d(k)\right)^{M-m}.
\end{split}
\end{equation}

It is relevant to point out that the classical optimizer employed in the \textbf{Training} step attempts to maximize the average fidelity of each final state $\ket{\psi}_i$---associated with a \textsc{training} dataset instance---with respect to its ground truth ($\ket{1}$ or $\ket{0}$ for fraudulent and non-fraudulent instances, respectively). Therefore, it is possible that not all final states associated with instances of the \textsc{testing} dataset will have a fidelity above $\tfrac{1}{2}$ with respect to their ground truth; this is a source of errors in the classification and, thus, represents a deterioration of the quality of service. Evaluation of the latter is performed based on a common figure of merit used in binary classification, the \textbf{f1-score}, which balances \textbf{precision} and \textbf{recall} -- see Appendix A Subsection 1 for details on binary classification metrics. Figure \ref{fig:appA_fid_raw} in Appendix A depicts the final distribution of fidelities for the \textsc{testing} dataset used in this study.

From the network performance perspective, one is interested in estimating the effect of channel quality on the classification metrics since it impacts the quality of service delivered by the proposed quantum network, analyzed in depth in the next subsection IV-B. In practice, the number of rounds $M$ can be increased to offset the decoherence introduced by the channel; however, this can impact not only the time it takes for the user to make a final classification decision but also parameters of the quantum communication platforms, such as the pump power of the client's entangled photon-pair source needed to deliver classifications at a required rate. In order to further evaluate the service quality as a function of the channel decoherence, the quantum circuit of the PB-SQC was implemented on Qiskit \cite{ibm2025qiskit}, where a depolarizing channel model was introduced allowing one to control the depolarizing error $\zeta_S$ acquired by the Server's qubit after RSP.

\begin{figure}[h]
    \centering
    \includegraphics[width=1\linewidth]{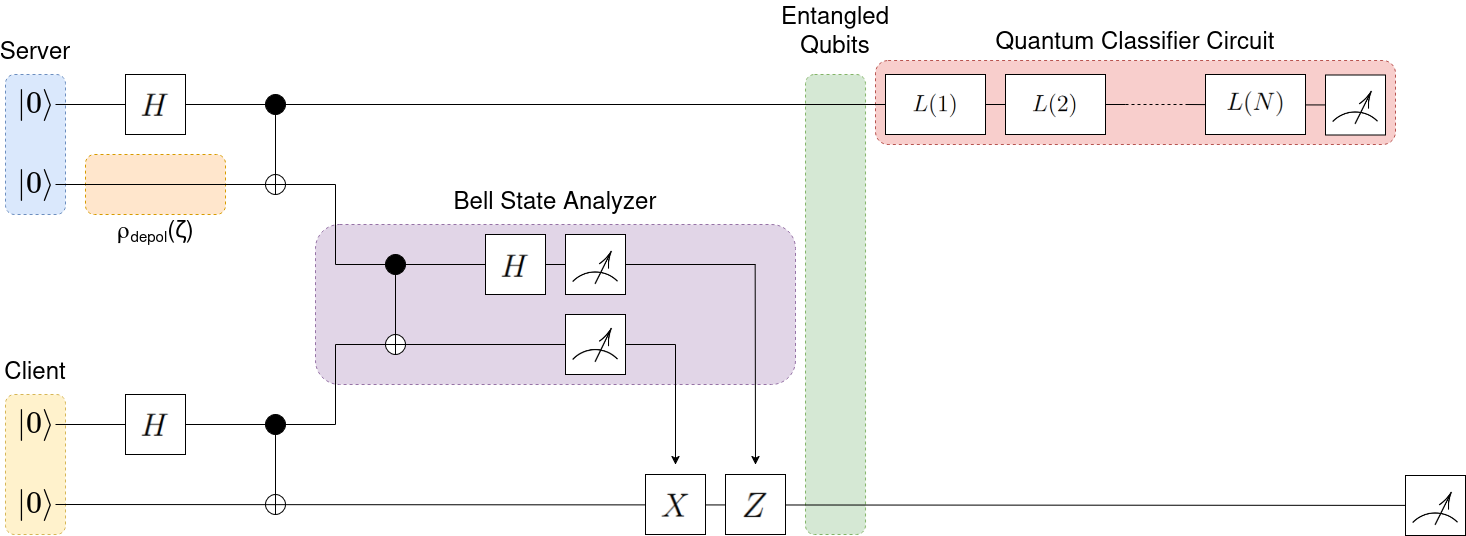}
    \caption{Final scheme of the quantum classifier circuit with entanglement swapping. Entanglement between qubits [0,1] and [2,3] is prepared via the application of a hadamard and CNOT gates at server and client, respectively. Qubits [1,2] are jointly measured at the intermediate node and the classical detection pattern information is transmitted to the Client, which performs RSP. The Server executes the padded rotations provided classicaly by the Client, measures on the computational basis, and transmits the classical result to the Client.}
    \label{fig:swap_qiskit}
\end{figure}

The quantum network architecture presented in Section III -- assuming an idealized scenario of deterministic sources of entanglement (light-matter and photon-pairs) -- is an entanglement swapping setup, which can be translated into the quantum circuit of Fig. \ref{fig:swap_qiskit}. There, RSP is performed on the Client's qubit (represented by the X and Z squares) after the intermediate node measurement (represented by the purple box) and the PB-SQC padded rotations are performed on the Server's qubit (represented by the pink box). The quantum channel quality can be tuned via the gate represented by $\rho_{depol}\left(\zeta\right)$. This quantum circuit was implemented following the protocol described in \textbf{Algorithm 1} on Qiskit.

\begin{figure}
    \centering
    \includegraphics[width=\linewidth]{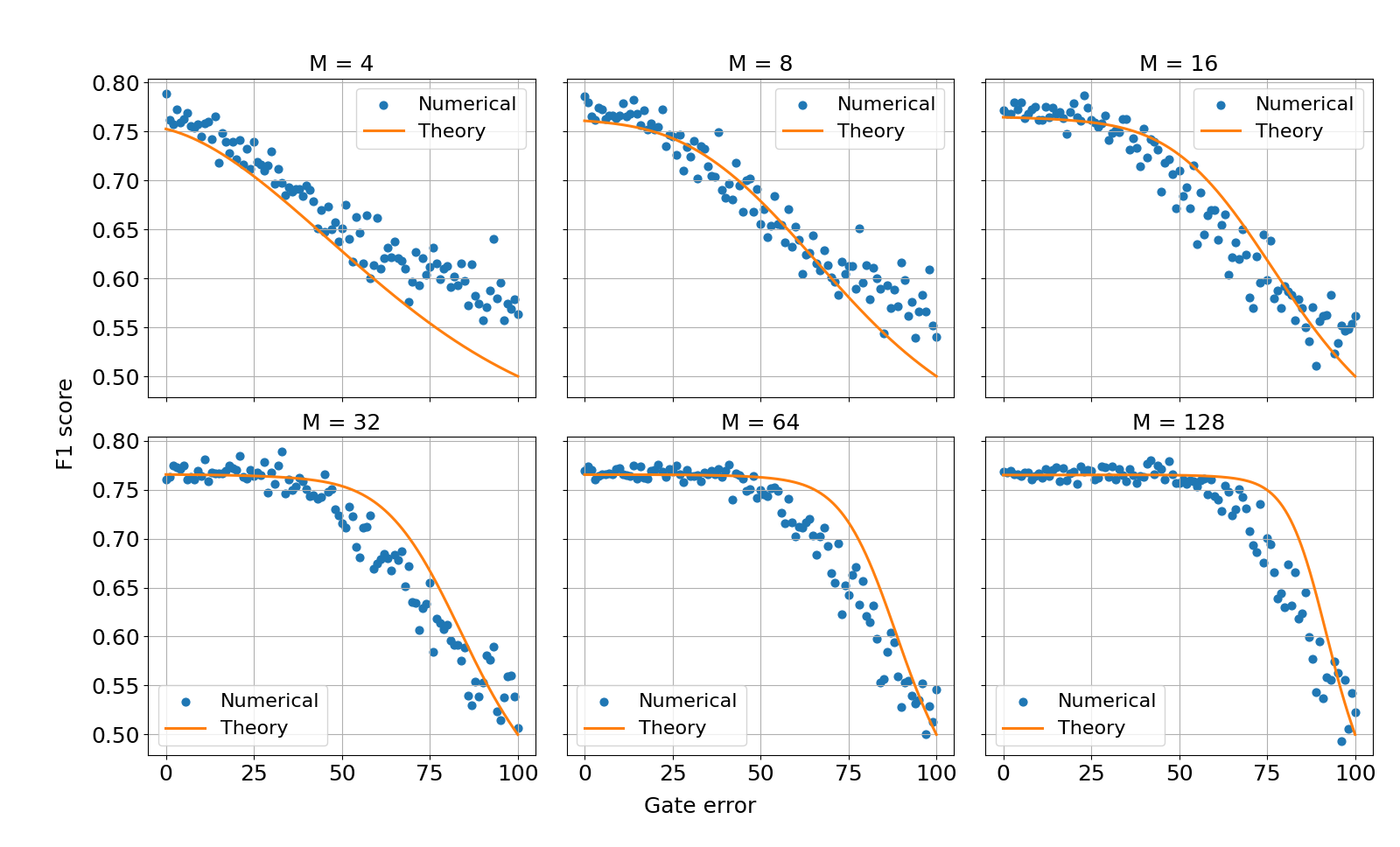}
    \caption{Performance of the PB-SQC protocol running on the Qiskit software framework assuming depolarizing channel errors. The higher \textbf{f1-score} values are recovered for larger M as the decoherence parameter $\zeta$ increases, which translates into longer time-to-service for the client.}
    \label{fig:qiskit_results}
\end{figure}

To validate the performance estimation analysis in presence of decoherence, the values of \textbf{f1-score} achievable by the PB-SQC protocol using the \textsc{testing} dataset are presented in Fig. \ref{fig:qiskit_results}. The numerical results acquired using the Qiskit setup are overlaid by the values calculated using Eqs. \ref{eq:prob_depol} and \ref{eq:P_afterM} (averaged over all instances $d$) given the values of $|\bra{k}\ket{\psi}_d|^2$ (Appendix A, subsection 2, Figure 1) of the \textsc{testing} dataset. The good agreement between the models lays the foundation for the network performance estimation of the next subsection. Furthermore, a closed form for $M$ is available: given $\zeta_S$ and a performance threshold $\delta$ (which can be set by the client and represents the acceptable ratio below the maximum classification performance), the required number of rounds $M$ is, by Hoeffding's inequality~\cite{hoeffding1963}:
\begin{equation}
\label{eq:m_repetitions}
    M\left(\zeta_S,\delta\right) = \tfrac{2}{\left(1-\zeta_S\right)^2}\log{\left(1/\delta\right)}.
\end{equation}
In summary, this analysis indicates that, with knowledge of the network, an operator can optimize performance and predict time-to-service for a given classification threshold imposed by the user.

\subsection{PB-SQC Performance over a Prototype Hybrid Quantum Network}
\label{sec:usecase_performance}

Finally, we estimate the performance of a fraud detection use-case based on PB-SQC implemented on our quantum network link prototype. We do this by combining the simulation results of the fraud detection algorithm in Sec. \ref{sec:fraud_results} with the performance of the quantum network link of Sec. \ref{sec:network_performance}. Our goal is to estimate the maximum time it would take for the client to obtain one classification with the highest possible classification performance (nominal \textbf{f1-score} with zero channel error).
Given an amount of depolarizing error $\zeta_T$ in each run, the classification protocol has to be repeated for $M(\zeta_T,\delta)$ times to yield the nominal \textbf{f1-score} value, with $M$ given by Eq. \ref{eq:m_repetitions}. It is important to note that this maximum value is given solely by the representativity of the training dataset and the performance of the classical optimizer used during supervised training. For each repetition, the client and server must successfully generate remote entanglement, which will be used to perform RSP. The finite entanglement fidelity will contribute to the depolarizing error $\zeta_T$. The total execution time is given by

\begin{align}
    t_\textrm{exec} = M(\zeta_T,\delta)t_\textrm{run}t_\textrm{ent}.
    \label{eq:execution_time}
\end{align}

Here, $t_\textrm{run} \approx n_\textrm{gates} \times t_\textrm{gate}$ is the execution time of a single protocol run, which is mainly determined by the number of gates to be applied by the server and the gate time, while $t_\textrm{ent}$ is the time required to successfully distribute entanglement between the server and client. Given a certain success probability of the entanglement generation, $p_\textrm{success}$, we can estimate the number of attempts $A$ needed to generate entanglement with $99\%$ probability:
\begin{align}
    A_{99\%} = \log(1-0.99)/\log(1-p_\textrm{success}). 
\end{align}

In our system, fidelity and success probability can be traded off by tuning the SPDC source brightness. Therefore, $A_{99\%}$ can be reduced if the protocol can accept a lower entanglement fidelity.  The entanglement distribution time can be expressed as
\begin{align}
    t_\textrm{ent} = A_{99\%} \times t_\textrm{single attempt},
\end{align}
where the time for a single entanglement attempt is dominated by the distance between client and server: the photon travels from the nodes to the heralding station, and the classical heralding signal has to be sent back.\\

The total depolarizing error $\zeta_T$ can be divided into two contributions: the RSP error $\zeta_S$ as a result of the finite fidelity of the entangled state generation, and the gate error for the server operations, $\zeta_G$. Note that modelling both as a fully depolarizing channel gives a worst-case scenario. We can generate client-server states with lower fidelity at a higher rate, but higher total depolarizing error requires more classification rounds $M$ to yield maximal \textbf{f1-score}. Thus, it is interesting to investigate if there is an optimum operation point for minimizing the time in Eq. \ref{eq:execution_time}. We do this by selecting the parameter regime of the network link that can provide the highest $p_\textrm{success}$ for a required fidelity $F_\textrm{req}$, which is determined given the gate error and total error $\zeta_T$ as
\begin{equation}
    F_\textrm{req} = 1 - \frac{3}{4} \bigg( \frac{\zeta_T - \zeta_G}{1-\zeta_G}\bigg).
\end{equation}

\begin{figure}
    \centering
    \includegraphics[width=\linewidth]{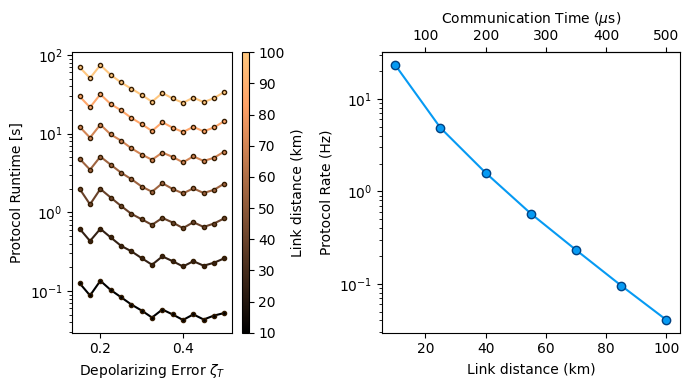}
    \caption{Simulation of the PB-SQC performance on a realistic quantum network link. (Right) Protocol runtime for different allowed depolarizing error $\zeta_T$ at different link distances. (Left) Optimal protocol runtime for each distance.}
    \label{fig:bsqc_performance}
\end{figure}

Figure \ref{fig:bsqc_performance} (right) shows the execution time as a function of the total error $\zeta_T$, for different total client-server distances. There is an optimal operation point when some error is allowed in the protocol. This is exactly because the faster entanglement generation at lower fidelity compensates for the required higher number of protocol rounds $M$, yielding an effectively faster run-time for the whole classification protocol. In the left plot, we show the optimal operation rate for different distances. The parameters used in the simulation are reported in Appendix \ref{app:link_performance}, Table \ref{tab:parameters_simulation}. These parameters reflect a realistic experimental implementation, as the efficiencies for the Rydberg system, QFC and quantum memory are compatible with near-term technical improvements over values that have already been experimentally realized. In particular, we consider $50\%$ and $60\%$ efficiency for the Rydberg photon generation and QFC, respectively. Values of $44\%$ \cite{yang2022deterministic} in-fiber efficiency were reported for Rydberg photon generation assisted by a low-finesse cavity, while QFC from rubidium wavelength to the telecom S-band (1522 nm) was reported with $64\%$ external efficiency before filtering \cite{Luo2025_entanglingquantummemories}. Cavity-enhanced quantum memories based on Pr:YSO crystals have demonstrated storage efficiency up to $62\%$ \cite{Duranti2024_efficientcavityassisted}, and up to $40\%$ using on-demand spin-wave storage \cite{Feldmann2025_cavityenhancedspinwave}. Recently, other platforms such as europium have shown single-photon storage efficiency up to $70\%$ \cite{meng2026_efficient}.

\section{Conclusion and Future Work}

In this work, we investigate application-driven protocols through which quantum communication networks can deliver practical utility and societal impact, with a focus on cryptographic advantage, which stands out as one of the core motivations behind the deployment of these networks.
We have introduced partially-blind single qubit classification (PB-SQC), a protocol for delegating the classification of dataset instances to a quantum computer, without the latter learning anything about the input or the output of the classification. This protocol has the potential of being realized in short-term networks and paves the way towards blind two qubit classification and its higher-qubit generalizations; in the latter case, distributed blind N-qubit classification can offer a computationally-complex application for long-term quantum networks

While single-qubit classification does provide noise-robustness, we only obtain verifiability of quantum computations with 2-qubit servers. Single-qubit classification can be suitable for trusted Servers on whom Clients trust to execute the designated computation but not to gain access to the data. Two-qubit classification, in contrast, does not require any trust to be deposited on the Server, as this can be verified by the Client, opening the door to commercialization of the service. Compared to other blind quantum machine learning protocols \cite{Fleur2025BQML}, the PB-SQC exhibits a low resource requirement (a single-qubit platform at both Server and Client): as emphasized in this paper, this comes from the quantum-classical learning feature of the protocol.

In the mathematical proofs of partial-blindness (and blindness in the 2-qubit case), perfect performance has been assumed, i.e., all errors arising during test rounds are assigned to the malicious behavior of the Server. The noise robustness of the PB-SQC is associated with classical error correction via repeated experiments to extract classification results from noisy data. Emphatically, while we prove blindness in this work, we provide no guarantees of composability: the property through which inclusion of the PB-SQC subroutine in a larger protocol would preserve blindness. This cryptographic task is left as an open question necessitating further study. 

Although we have successfully shown how using of Haar-random unitaries, $1$-designs, and remotely prepared states in the PB-SQC protocol can be used to achieve partial blindness, it is still an open question whether a more efficient scheme can be achieved including in the multi-qubit generalizations, requiring a different execution of the network. Also left as a future research direction is the inclusion of multi-qubit extensions considering the multiplexing capacity of the quantum communication platforms considered in the network architecture; there, special attention shall be given to qu\textbf{d}its and multi-qubit encoding onto higher-dimension quantum state. Finally, this work paves the way for near-term experiment in blind quantum machine learning with the potential to address the question of usefulness and societal impact expected from quantum information networks, with scaling up to multiple qubits providing an opportunity to unify quantum speedups and quantum privacy advantages with practical utility. 

\section*{Acknowledgments}
The authors are forever indebted to the QuTech Band, responsible for bringing this collaboration together. A. W. and G. C. A. acknowledge the NITeQ PUC-Rio group, especially J. M. Neto, B. Povoa, and M. dos Magos. G. M. L. and G. C. A. acknowledge W. Koz\l{}owski for valuable input and support. This project has received funding from the European Union's Horizon Europe research and innovation programme under grant agreement No. 101102140. TzBP gratefully acknowledges support from the Quantum Software Consortium Ada Lovelace
Fellowship. The dataset used for model training was provided by the fintech company \textit{Stone Pagamentos S.A.}

\section*{Data Availability}
The code used for simulating the PB-SQC and the network link hardware performances is available from the corresponding authors upon request.

\newpage
\hspace{1pt}
\newpage

\appendix{\section*{\textbf{Appendix A - Fraud Detection and the PB-SQC}}}
\label{app:fraud_detection}

\subsection{Classification Metrics}

We use a dataset \cite{neto2026fraud} provided by a Brazilian fintech, composed of the metadata of financial transactions (name, civil identifier number, time, location, etc.) between December 2023 and March 2024. The goal is to label the past data based on operator knowledge as fraudulent and non-fraudulent so that \textit{supervised learning} can be implemented allowing for future classification of transactions. To evaluate the model's performance during training, a loss-function can be used to indicate how well the model's prediction is aligned with the classification results. Generally, the number of fraudulent cases is overwhelmingly smaller than that of non-fraudulent ones in the complete dataset, which creates a bias of the loss-function that prevents the model from correctly identifying the fraudulent cases and, thus, achieve its goal. Therefore, the dataset is pre-processed to reflect an equal distribution of instances that belong to the two classes (\textbf{Dataset Preparation} step in the Main Text. After going through \textbf{Training} the classification performance is evaluated with a different dataset (\textsc{training} and \textsc{testing}, respectively), for which the following definitions are relevant in the context of a binary classification process and are used to calculate the \textbf{f1-score} metric utilized in Section V of the Main Text.
\begin{itemize}
\item [\textbullet]\textbf{True positive (TP):} a positive instance that was correctly classified as positive by the model.
\item [\textbullet]\textbf{True negative (TN):} a negative instance that was correctly classified as negative by the model.
\item [\textbullet]\textbf{False positive (FP):} a negative instance that was incorrectly classified as positive by the model.
\item [\textbullet]\textbf{False negative (FN):} a positive instance that was incorrectly classified as negative by the model.
\end{itemize}
From these definitions, the ones below follow:
\begin{itemize}
\item [\textbullet]\textbf{Precision:} The ratio of true positives to the total predicted positives. This metric indicates how many of the predicted fraudulent cases were actually fraud.
\[
\text{Precision} = \frac{\#TP}{\#TP + \#FP}
\]
\item [\textbullet]\textbf{Recall}: The ratio of true positives to the actual positives. It measures the model's ability to identify actual fraud cases.
\[
\text{Recall} = \frac{\#TP}{\#TP + \#FN}
\]
\item [\textbullet]\textbf{F1 Score:} The harmonic mean of precision and recall, providing a single score that balances equally both metrics.
\begin{equation}
\text{F1 Score} = 2 \times \frac{\text{Precision} \times \text{Recall}}{\text{Precision} + \text{Recall}}
\label{eq:f1score}    
\end{equation}
\end{itemize}

By employing these metrics, it is possible to get a clearer understanding of the model's performance in terms of its ability to correctly classify fraudulent and non-fraudulent transactions. The Recall metric, in particular, is of interest to credit card transaction operators because it is often preferable to mistakenly classify a non-fraudulent transaction as a fraud than letting a fraud go unnoticed, which can potentially cause significant financial damage.

\subsection{Training}

As detailed in Ref. \cite{perez2020data}, it follows from the Universal Approximation Theorem and the Universal Quantum Circuit Approximation Theorem that a quantum circuit can approximate a continuous classification function. In other words, it is possible to reconstruct any continuous function that represents a neural network with a single hidden layer of $N$ neurons. In turn, this $N$-neuron hidden layer can be implemented---via data-reuploading---on a single-qubit(Fig. \ref{fig:BSQC} of the Main Text) such that:
\begin{equation}
\mathcal{U} (\vec{\phi}, \vec{x}) \equiv U (\vec{\phi}_{N}) U (\vec{x}) \dots U (\vec{\phi}_{1}) U (\vec{x})
\end{equation}
and, thus,
\begin{equation}
\ket{\psi} = \mathcal{U} (\vec{\phi}, \vec{x}) \ket{0},
\end{equation}
where the vectors \( \vec{x} \) and \( \vec{\phi} \) represent the input data and the training biases, respectively, \( N \) is the number of processing units in the circuit (equivalent to the number of neurons in the hidden layer of a single-layered neural network), \( \ket{0} \) is the initial state of the circuit and \( \ket{\psi} \) is its final state.

\begin{figure*}
\centering
\includegraphics[width=0.45\textwidth]{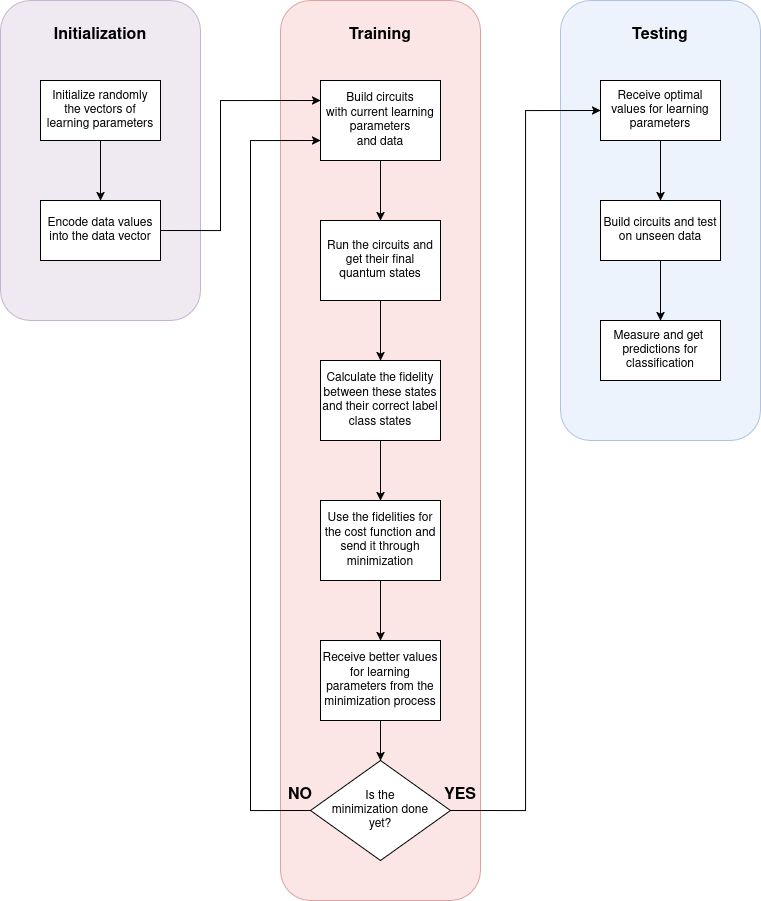}
\caption{Flowchart of the model's training.}
\label{fig:flow}
\end{figure*}

During training, a classical optimizer solves the problem of minimizing the \textbf{fidelity cost function} defined
\begin{equation}
\begin{split}
&\chi^2_f(\vec{\theta}, \vec{w}) = \\ 
&\sum_{\lambda=1}^{\Lambda} \left( 1 - \left| \braket{\tilde{\psi}_d | \psi(\vec{\theta}, \vec{w}, \vec{x}_\lambda)} \right|^2 \right),
\end{split}
\end{equation}
where $\ket{\tilde{\psi}}_s$ corresponds to the ground truth of the final state associated with the $d$-th instance of the \textsc{training} dataset and $\left(\Lambda=2500\right)$ is this dataset's size. The complete flowchart of the classification performance evaluation (including the three steps described in Section V of the Main Text) is presented in Fig. \ref{fig:flow}.

\begin{figure}
    \centering
    \includegraphics[width=\linewidth]{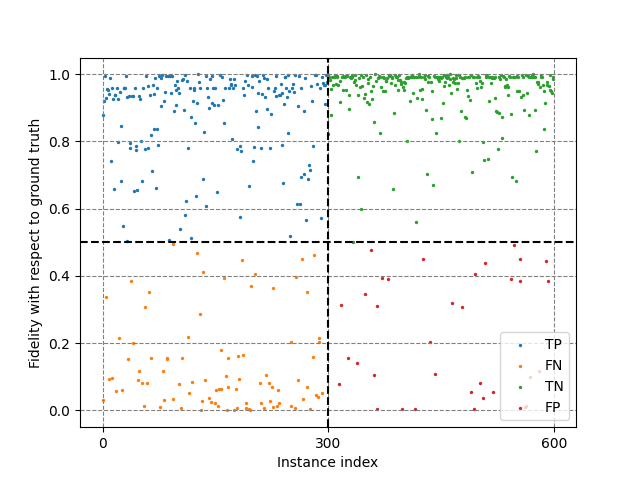}
    \caption{Fidelities of states $\ket{\psi}_s$ associated with instances of the \textsc{testing} dataset with respect to the ground truth at zero channel decoherence. Instance indices 1 to 300 are fraudulent and 301 to 600 are non-fraudulent. Extracting this result is only possible because the labels of each instance are available during the \textbf{Evaluation} step of the \textit{supervised learning} method.}
    \label{fig:appA_fid_raw}
\end{figure}

In possession of the training biases, it is possible, given the labels of the \textsc{testing} dataset, to determine each final state's fidelity with respect to the ground truth, presented in Fig. \ref{fig:appA_fid_raw}, which can then be inserted as $|\bra{k}\ket{\psi}_i|^2$ into Eq. \ref{eq:prob_depol}. Furthermore, as the probability $P_d\left(M,k\right)$ tends to 1 (0) when the value of $P_d\left(k\right)$ is above (below) $\tfrac{1}{2}$, one can use the results of Fig. \ref{fig:appA_fid_raw} to define instances that correspond to true/false positives/negatives in the zero-decoherence case.

\appendix{\section*{\textbf{Appendix B - Calculation of the Quantum Network Link Performance}}}
\label{app:link_performance}

\subsection{Theoretical model and methods}

Let us estimate the expected entanglement generation performance of the quantum link. Starting from the quantum states defined in equations \ref{eq:psi_spdc} and \ref{eq:psi_server}, we can write the joint state of the server and client before the mid-point station as
\begin{align}
    \ket{\psi_\textrm{0}} = \ket{\psi_\textrm{server}}\otimes\ket{\psi_\textrm{SPDC}}.
\end{align}
We introduce losses by applying a beam splitter operation to each photonic mode:
\begin{align}
    \hat{a}_i^\dagger \to \sqrt{\eta_{l}}\hat{a}_i^\dagger + \sqrt{1-\eta_{l}}\hat{a}_{i,\textrm{loss}}^\dagger
\end{align}
where $i$ indicates the photonic mode, $\eta_l$ is the photon transmission efficiency for each component of the quantum link ($m = \textrm{server, SPDC. QM}$). The modes $\hat{a}_{i,\textrm{loss}}$ are traced out after heralding. This is implemented in practice using the unitary operator
\begin{equation}
\begin{split}
    U_{i,\textrm{loss}} &= e^{i\theta_{l}(\hat{a}_i^\dagger\hat{a}_{i,\textrm{loss}} + \hat{a}_i\hat{a}_{i,\textrm{loss}}^\dagger)}\\
    \theta_l &= \arctan\left( \sqrt{\frac{1-\eta_{l}}{\eta_{l}}}\right).
\end{split}
\end{equation}

We account for the possible detection outcomes, including single-photon and two-photon detection events, following an approach similar to \cite{Hermans2023}. For simplicity, we consider only one heralding pattern out of the four that can happen, namely, a "click" is recorded in the same detector $\left(D_1\right)$ for each of the two time-bins. This heralds the state in Eq. \ref{eq:psi_entangled}. The other valid heralding patterns are: a "click" in detector B for each time-bin, one click in $D_1$ $\left(D_2\right)$ at the Early time-bin and in $D_2$ $\left(D_1\right)$ at the Late time-bin. The heralding patterns where the clicks happen in two different detectors result in a different Bell state, and can be correctly used for the protocol, as long as the client has access to the heralding pattern.\\
The ideal case, where only a single photon per time-bin reaches the detector input, leads to the projective Kraus operator
\begin{align}
    P_{1ph} = \bra{vac}\hat{a}_{D_1,e}\hat{a}_{D_1,\ell}.
    \label{eq:project_sph}
\end{align}

We make use of the beam-splitter relations
\begin{equation}
\hat{a}_{D_1}^\dagger\to\frac{1}{\sqrt{2}}(\hat{a}_\textrm{server} + \hat{a}_\textrm{SPDC})    
\end{equation}
to rewrite this expression as a function of the photonic modes in $\ket{\psi_0}$ and calculate the resulting density matrix
\begin{align}
    \rho_{1ph} = P_{1ph} \ket{\psi_0}\bra{\psi_0}P_{1ph}^\dagger.
    \label{eq:rho_sph}
\end{align}
In the ideal case, where no loss and no multiphoton components are present, this heralds the ideal entangled state of Eq. \ref{eq:psi_entangled}
\begin{align}
    \rho_\textrm{ent} =  \ket{\psi_\textrm{ent}}\bra{\psi_\textrm{ent}}.
\end{align}

In most realistic experimental implementations, the detectors cannot resolve the number of photons. Therefore, detection of two-photon states will be indistinguishable from the ideal one-photon case, and will contribute to the final heralded state as if it were a single-photon detection. This reduces the fidelity of the final state. We calculate the density matrix associated with the case where detector A receives two photons in Early and one in Late as
\begin{equation}
    \begin{split}
        P_{2ph,E} &= \bra{vac}(\hat{a}_{D_1,e})^2\hat{a}_{D_1,\ell}\\
        \rho_{2ph} &= P_{2ph}\ket{\psi_0}\bra{\psi_0}P_{2ph}^\dagger,
    \end{split}
\end{equation}
and similarly for the case of two photons in Late and one in Early, $\rho_{2ph,\ell}$.

Next, we compute the case where one photon is lost, and two are heralded. Photon loss will not conserve the coherence of the state, thus we can compute all possible cases separately and sum the density matrices incoherently. The case for a lost photon in mode $i$ is given by
\begin{equation}
    \begin{split}
        P_{i,\textrm{loss}} &= \bra{vac}\hat{a}_{D_1,e}\hat{a}_{D_1,\ell}\hat{a_{i,\textrm{loss}}}\\
        \rho_{i,\textrm{loss}} &= P_{i,\textrm{loss}}\ket{\psi_0}\bra{\psi_0}P_{i,\textrm{loss}}^\dagger.
    \end{split}
\end{equation}

The total heralded density matrix is
\begin{align}
    \rho_\textrm{herald} = \rho_{1ph} + \sum_{k=e,\ell}\rho_{2ph,k} + \sum_{i}\rho_{i,\textrm{loss}},
    \label{eq:rho_herald}
\end{align}
where the sum over $i$ covers all photonic modes. We account for dark counts at the BSM detectors, which can result in heralding errors. There is a finite probability $p_{DC}^{x,Y}$ that a noise count is detected at detector $x$ during the detection window of time-bin $Y$. When this happens, for example, on detector $D_1$ and Early time-bin, the heralding pattern of Eq. \ref{eq:rho_sph} becomes:
\begin{align}
    P_{1ph}\hat{a}^\dagger_{D_1,e}\ket{\psi_0} = \bra{0}\hat{a}_{D_1,\ell}\ket{\psi_0}.
\end{align}

We compute all combinations of one dark counts in one time-bin for the single- and two-photon heralding events, including the ones with lost photons, and we sum them into a noise density matrix $\rho_\textrm{noise}$. We neglect the case where two dark counts are detected, since the probability is negligible. Finally, we combine the result of noise-affected heralding with the density matrix of Eq. \ref{eq:rho_herald}:
\begin{align}
    \rho_\textrm{final} = (1-p_{DC}) \rho_\textrm{herald} + p_{DC}\rho_\textrm{noise}.
\end{align}

We calculate the fidelity and the success probability of obtaining the target state
\begin{align}
    \begin{split}
        F = \left( \Tr{ \sqrt{ \sqrt{\tilde{\rho}}\rho_\textrm{ent}\sqrt{\tilde{\rho}}}}\right)\\
        p_\textrm{success} = \Tr{\rho_\textrm{final}},
    \end{split}
    \label{eq:ent_metrics}
\end{align}
where 
\begin{align}
\begin{split}
    \tilde{\rho} = \frac{1}{\Tr(\rho_\textrm{final})}\rho_\textrm{final}
\end{split}
\end{align}
is the normalized final density matrix. Note that here we only considered one out of four possible detection patterns (one click in each Early and Late, in all possible combinations of detectors), all of which herald a maximally entangled state. Therefore, the actual success probability of the protocol is four times that of Eq. \ref{eq:ent_metrics}.

\begin{figure}
    \centering
    \includegraphics[width=\linewidth]{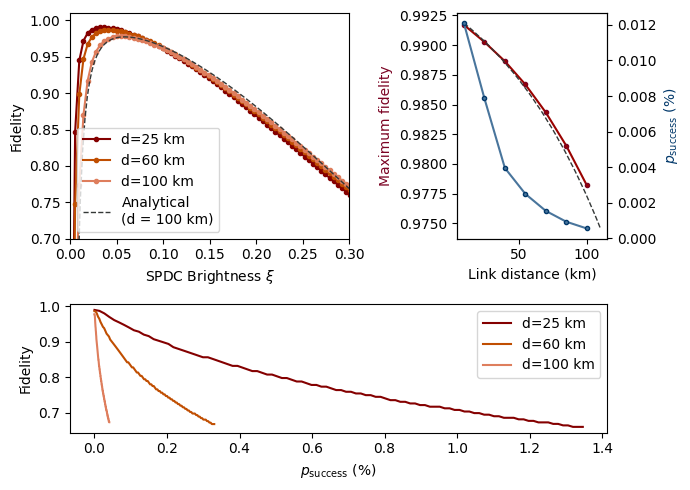}
    \caption{Performance of the entanglement generation in the hybrid client-server link. (Top left) Sweep of the SPDC brightness $\xi$, for three example distances. (Top right) Maximum achievable fidelity and corresponding success probability for different link distances. The dashed black line is the analytical approximation of Eq. \ref{eq:analytical_fidelity}. (Bottom) Rate-fidelity tradeoff for three link distances.  }
    \label{fig:link_performance}
\end{figure}

We compute the metrics in Eq. \ref{eq:ent_metrics} numerically using QuTip \cite{qutip5}. We set the efficiencies $\eta_m$ to incorporate setup losses. In particular, we consider losses due to finite Rydberg photon generation and collection efficiency and QFC for the server, efficiency of the QM, and loss due to fiber transmission in the SPDC and Server modes (assuming $0.2$ dB/km attenuation). The parameters used are reported in Table \ref{tab:parameters_simulation}. Varying the brightness parameter $\xi$ of the SPDC enables finding a tradeoff between rate and fidelity. Figure \ref{fig:link_performance} shows the maximum fidelity achievable for different distances between the nodes and the midpoint station, and a rate-fidelity curve for different distances.

\subsection{Impact of noise and multiphoton components in the Rydberg Superatom}

In this model, we have considered the Rydberg superatom as a perfect single-photon emitter. This is the case for an atomic cloud that is smaller than the Rydberg blockade radius, where strictly one Rydberg excitation is allowed. Such a condition is experimentally achievable, but challenging to obtain, and in realistic experiments, the atomic cloud can allow multiple excitations, decreasing the purity of the single photons generated. In addition, QFC can introduce noise due to, e.g. spontaneous conversion processes enabled by the strong pump. This noise can be strongly suppressed by engineering the QFC interaction and by narrow spectral filtering. For these reasons, in our model, we neglect the imperfections in the Rydberg and QFC system. Here, we estimate the conditions under which this assumption is valid, and the error given by the finite single-photon purity of the server is subdominant compared to that caused by the multiphoton components from the SPDC source.\\

In our protocol, we herald over two detection events (clicks), one in each time-bin. This heralds the correct entangled state in the case where one click comes from each node (Rydberg or SPDC). This happens with probability:
\begin{align}
    P_\textrm{correct} \sim p_R p_S,
\end{align}
where $p_{R(S)}$ indicates the probability of a Rydberg (SPDC) click. Note that $p_S \approx \xi$ We have three main causes for erroneous heralding. The first is when two clicks are caused solely by the SPDC, and no photon from the Rydberg. Assuming low SPDC brightness, this happens with probability
\begin{align}
    P_\textrm{multiphoton} \sim p_S^2 + O(p_S^3).
\end{align}

Assume we have noise in the Rydberg channel which is distinguishable (i.e. that does not interfere at the beam splitter) and uncorrelated with the atomic state, and gives a click with probability $p_N$. The source of this noise could be, for example, spurious excitation due to imperfect Rydberg blockade, noise coming from QFC, or dark counts at the detectors. We can have a false heralding event when one click is given by noise and the other is given by either the SPDC or the Rydberg. This happens with probability

\begin{align}
    P_\textrm{Noise} \sim p_Np_S + p_Np_R.
\end{align}

By assuming orthogonality of the resulting states, we can estimate the fidelity after heralding

\begin{align}
\label{eq:analytical_fidelity}
  \mathcal{F} \approx \frac{P_\textrm{correct}}{P_\textrm{correct} + P_\textrm{multiphoton} + P_\textrm{noise}} \\\approx \frac{1}{1 + p_S/p_R + p_N/p_S + p_N/p_R}.
\end{align}

We can write the condition that the noise effect is negligible compared to the multiphoton error as
\begin{align}
    \frac{p_N}{p_S} + \frac{p_N}{p_R} \ll \frac{p_S}{p_R} \to p_N \ll \frac{p_S^2}{p_S+ p_R}.
\end{align}

We can use this model to estimate the requirements on the purity of the single photons coming from the Rydberg server. Consider that the $g^{(2)}_R(0)$ of the Rydberg system, including Poissonian uncorrelated noise, can be written as
\begin{align}
    g^{(2)}_R(0) = 1 - \left(\frac{p_R}{p_N + p_R}\right)^2.
\end{align}

Solving for $p_N$, and assuming $g^{(2)}<<1$, this leads to the condition
\begin{align}
    g^{(2)}_R \ll \frac{2p_S^2}{p_Sp_R + p_R^2}.
\end{align}
Note that, in general, in experiments we have that $p_S \ll p_R$. In this case, this simplifies to
\begin{align}
    g^{(2)}_R \ll 2\left(\frac{p_S}{p_R}\right)^2.
\end{align}

\subsection{Parameters for Use Case Performance Simulation}

\begin{table}[h!]
    \centering
    \begin{tabular}{c|c}
         Parameter & Value \\
         \hline
         $n_\textrm{gates}$ & 18 \\
         $t_\textrm{gate} $ & 100 ns \\
         $t_\textrm{single attempt}$ & $d_\textrm{km} / (2\times10^5 \frac{km}{s})$ \\
         $\zeta_G$ & 0.1 \\
         $p_{DC}$ & $10 \textrm{Hz} \times 200 \textrm{ns}$\\
         $\eta_\textrm{QFC}$ & 0.6\\
         $\eta_\textrm{Rydberg}$ & 0.5\\
         $\eta_\textrm{QM}$ & 0.6\\
         \hline
    \end{tabular}
    \caption{Parameters used for the simulation in Fig. \ref{fig:bsqc_performance}. The efficiencies $\eta$ reported are to account for the frequency conversion (QFC) of the server photon, collection efficiency of the Rydberg photon emission, and overall quantum memory (QM) efficiency.}
    \label{tab:parameters_simulation}
\end{table}

For the simulations in Sec. \ref{sec:usecase_performance}, we use the parameters in Table \ref{tab:parameters_simulation}. Note that, in the timings of the entanglement generation and PB-SQC protocol implementation, we do not take into account the reduced duty cycle of the Rydberg-based server due to atom cooling and trapping. Thus, the reported protocol rate can be understood as an upper bound.\\

\bibliography{BSQC.bib}


\end{document}